\documentclass[final,11pt,3p]{elsarticle}
\usepackage{hyperref}

\usepackage{float}
\newfloat{chart}{htbp}{loc}
\floatname{chart}{Listing}

\usepackage{tabularx}
\journal{---}

\usepackage{moreverb}
\usepackage{siunitx}
\usepackage{tabularx}
\usepackage{amsmath}
\usepackage{amssymb}
\usepackage[acronym]{glossaries}
\usepackage{graphicx} 
\usepackage[process=auto,crop=pdfcrop]{pstool}
\usepackage[singlelinecheck=true]{subfig}
\usepackage{mathtools}
\usepackage{xcolor}
\usepackage{booktabs}
\usepackage{array}
\usepackage{setspace}
\usepackage{multirow}
\usepackage{etoolbox} 
\usepackage{xcolor}
\usepackage{lmodern}
\usepackage{listings}
\usepackage{fancyvrb}
\usepackage{caption}
\usepackage{verbatimbox}
\usepackage{tcolorbox}
\usepackage{etoolbox}
\usepackage[obeyspaces]{xurl}

\newcommand*\linenomathpatch[1]{%
  \cspreto{#1}{\linenomath}%
  \cspreto{#1*}{\linenomath}%
  \cspreto{end#1}{\endlinenomath}%
  \cspreto{end#1*}{\endlinenomath}%
}

\DeclareCaptionLabelFormat{cont}{#1˜#2\alph{ContinuedFloat}}

\linenomathpatch{equation}
\linenomathpatch{gather}
\linenomathpatch{multline}
\linenomathpatch{align}
\linenomathpatch{alignat}
\linenomathpatch{flalign}

\definecolor{darkgreen}{rgb}{0.,0.5,0.}
\definecolor{mybg}{rgb}{0.85,0.85,0.85}

\definecolor{codegreen}{rgb}{0,0.6,0}
\definecolor{codegray}{rgb}{0.5,0.5,0.5}
\definecolor{codepurple}{rgb}{0.58,0,0.82}
\definecolor{backcolour}{rgb}{0.95,0.95,0.92}

\lstdefinestyle{mystyle}{
    backgroundcolor=\color{mybg},   
    keywordstyle=\color{codegreen},
    basicstyle=\ttfamily\footnotesize,
    breakatwhitespace=false,         
    breaklines=true,                 
    showspaces=false,                
    showstringspaces=false,
    showtabs=false,                  
}

\begin{document}

\begin{frontmatter}

\title{
High-speed turbulent flows towards the exascale:\\ STREAmS-2 porting and performance}
\author[1]{Srikanth Sathyanarayana}
\address[1]{Department of Mechanical and Aerospace Engineering, Sapienza University of Rome, via Eudossiana 18, 00184 Rome, Italy}
\ead{srikanth.cs@uniroma1.it}

\author[1]{Matteo Bernardini\corref{mycorrespondingauthor}}
\cortext[mycorrespondingauthor]{Corresponding author}
\ead{matteo.bernardini@uniroma1.it}

\author[2]{Davide Modesti}
\address[2]{Aerodynamics Group, Faculty of Aerospace Engineering, Delft University of Technology, Kluyverweg 2, 2629 HS Delft, The Netherlands}
\ead{d.modesti@tudelft.nl}

\author[1]{\\Sergio Pirozzoli}
\ead{sergio.pirozzoli@uniroma1.it}

\author[3]{Francesco Salvadore}
\address[3]{HPC Department, Cineca, Rome office, via dei Tizii 6/B Rome, Italy}
\ead{f.salvadore@cineca.it}

\begin{abstract}
Exascale High Performance Computing (HPC) represents a tremendous opportunity to push the boundaries of Computational Fluid Dynamics (CFD), but despite the consolidated trend towards the use of Graphics Processing Units (GPUs), programmability is still an issue. STREAmS-2 (Bernardini \textit{et al.} Comput. Phys. Commun. 285 (2023) 108644) is a compressible solver for canonical wall-bounded turbulent flows capable of harvesting the potential of NVIDIA GPUs. Here we extend the already available CUDA Fortran backend with a novel HIPFort backend targeting AMD GPU architectures. The main implementation strategies are discussed along with a novel Python tool that can generate the HIPFort and CPU code versions allowing developers to focus their attention only on the CUDA Fortran backend. Single GPU performance is analysed focusing on NVIDIA A100 and AMD MI250x cards which are currently at the core of several HPC clusters. The gap between peak GPU performance and STREAmS-2 performance is found to be generally smaller for NVIDIA cards. Roofline analysis allows tracing this behavior to unexpectedly different computational intensities of the same kernel using the two cards. Additional single-GPU comparisons are performed to assess the impact of grid size, number of parallelised loops, thread masking and thread divergence. Parallel performance is measured on the two largest EuroHPC pre-exascale systems, LUMI (AMD GPUs) and Leonardo (NVIDIA GPUs). Strong scalability reveals more than 80\% efficiency up to 16 nodes for Leonardo and up to 32 for LUMI. Weak scalability shows an impressive efficiency of over 95\% up to the maximum number of nodes tested (256 for LUMI and 512 for Leonardo). This analysis shows that STREAmS-2 is the perfect candidate to fully exploit the power of current pre-exascale HPC systems in Europe, allowing users to simulate flows with over a trillion mesh points, thus reducing the gap between the Reynolds numbers achievable in high-fidelity simulations and those of real engineering applications.
\end{abstract}


\end{frontmatter}

\section{Introduction} \label{sec:introduction}

Recent computer architectures have become increasingly heterogeneous with the use of multi-node systems equipped with high-performance accelerators. Graphical Processing Units (GPUs) have been at the forefront of these changes and are the driving architecture on the current path to exascale computing. Currently, seven of the top ten supercomputers use GPUs for acceleration \cite{top500}. Much of the success has been attributed to optimised architectures that deliver high performance per watt. However, programmability is generally a major drawback, but programming efforts over the past few years, recent standardised paradigms and performance libraries have alleviated this challenge. To this end, NVIDIA is the most widely used GPU with its popular CUDA programming framework \cite{cuda}. In recent years, several legacy computational fluid dynamics (CFD) codes have been ported to NVIDIA architectures
and new ones have been developed to target this backend. AFiD \cite{zhu2018afid} and CANS \cite{costa2021gpu} are popular open source incompressible Direct Numerical Simulations (DNS) codes. AFiD uses a CUDA Fortran \cite{cuda-fortran} approach, while CaNS mostly uses OpenACC \cite{openacc}. For compressible flows, PyFR \cite{witherden2014pyfr} and ZEFR \cite{romero2020zefr} are unstructured based flow solvers using CUDA for GPU acceleration. More recently, URANOS \cite{de2023gpu} has been developed using OpenACC directives and is capable of running in DNS, Large Eddy Simulations (LES) and Wall-Modelled LES (WMLES) modes. The unstructured flow solver charLES also uses NVIDIA GPUs~\citep{bres_22}. To date, almost every approach to CFD simulation and modelling has been attempted using GPUs \cite{kuznik2010lbm}\cite{zhang2020novel}.

An important development in recent years has been the emergence of new GPU vendors with different hardware and programming environments. This has created serious problems for developers, forcing them to maintain multiple versions of the same code targeting different GPU architectures. This is also reflected in the planned exascale/pre-exascale supercomputers in the US and EU. Frontier at Oak Ridge National Laboratory, El Capitan at LLNL and LUMI at CSC will all use AMD GPUs, while Aurora at Argonne National Laboratory will use Intel GPUs and Leonardo at Cineca will use NVIDIA GPUs. This has sparked discussion about the use of portable programming solutions to tackle all the latest GPUs, regardless of their architecture specifications. Neko, part of the Nek5000 family, a spectral element based CFD code, has recently released its portable version \cite{jansson2021neko}. They use multi-level abstractions using abstract Fortran types, which facilitates the implementation of hardware-specific backends. PyFR uses Python's built-in domain-specific language tool, derived from the Mako templating engine, to generate backends for different software paradigms. Both Neko and PyFR have the ability to target vendor specific software such as CUDA or HIP \cite{hip} and general purpose frameworks such as OpenCL \cite{opencl}.  

Over the years, our compressible flow solver STREAmS has responded to the changing nature of High Performance Computing (HPC) and recognised the potential of GPUs. An earlier version of STREAmS was ported to a single GPU to target the NVIDIA Fermi architecture \cite{salvadore2013gpu}. This was followed by a full port to CUDA Fortran (hereafter called STREAmS-1) \cite{bernardini2021streams}. More recently, recognising the need to extend STREAmS-1 to become a truly portable solver, a complete refactoring called STREAmS-2 was developed to support multi-backend and multi-equation applications in an object-oriented manner \cite{bernardini2023streams}. This continuous evolution of the solver is mainly due to its maturity resulting from a history of more than 20 years dedicated to the study of compressible wall-bounded flows. Several seminal DNS studies on wall-bounded canonical flows such as supersonic and hypersonic boundary layers \cite{pirozzoli_bernardini_2011}\cite{pirozzoli_probing}\cite{cogo_salvadore_picano_bernardini_2022}, shock boundary layer interactions \cite{bernardini_della_posta_salvadore_martelli_2023}, supersonic internal flows \cite{modesti2016reynolds}. In addition, the solver has been extended to study some of the more challenging problems such as the supersonic roughness-induced transition \cite{bernardini2012compressibility}\cite{roughness_bernardini} and the effects of distributed surface roughness \cite{modesti_sathyanarayana_salvadore_bernardini_2022}.

In this work we extend the capabilities of STREAmS-2 by developing our first GPU portable version, which is designed to run on both NVIDIA and AMD GPUs in addition to the traditional CPU version. We begin with a brief background on the governing equations and numerical methods that form the basis of the algorithms used in the solver. We then provide a detailed description of the strategies used to develop the portable variant of STREAmS-2. Finally, we present results based on single-GPU performance on different architectures, followed by scalability results based on multi-node, multi-GPU HPC clusters. 

\section{Numerical Methods} \label{sec:numerics}

STREAmS-2 solves the compressible Navier-Stokes equations for an ideal gas in Cartesian coordinates using a finite difference discretisation. The nonlinear terms are treated using a hybrid discretisation that switches between a central scheme in smooth flow regions and a shock-capturing scheme in shocked regions. Numerical stability in smooth flow regions is achieved by using the built-in anti-aliasing properties of the skew-symmetric form of the convective terms, which are cast in terms of numerical fluxes to allow easy hybridisation with the shock capturing scheme~\cite{pirozzoli_10}. This formulation guarantees discrete conservation of kinetic energy in the inviscid incompressible limit. Optionally, one can also choose the latest KEEP-n scheme~\cite{tamaki_22}, which also guarantees local entropy conservation for inviscid smooth flows.

In the vicinity of discontinuities, a weighted essentially non-oscillatory (WENO) reconstruction is used to obtain the characteristic fluxes at the cell faces, which are then projected onto the right eigenvectors of the Euler equations. The switch between the two discretisations is controlled by a modified version of the classical Ducros shock sensor, which activates the shock capturing algorithm only close to discontinuities~\cite{bernardini2021streams}. The implementation is general enough to allow an arbitrary order of accuracy, although at the moment discretisations up to $8^{th}$ and $7^{th}$ order are available for the central scheme and WENO reconstruction respectively. Viscous terms are expanded in Laplacian form and discretised with central finite difference formulae up to an eighth order of accuracy. For time integration we use a third-order low storage Runge-Kutta scheme.

A feature of STREAmS-2 is that both calorically and thermally perfect gases can be simulated, whereby the dependency of the specific heats on temperature is accounted for using NASA polynomials~\cite{mcbride_02}.

Apart from this last feature, the numerical approach described here has been used by the group for two decades to study compressible wall-bounded flows. The main novelty of STREAmS-2 is that the legacy solver has been rewritten using an object-oriented flavour to allow future extensions and porting to different backends, as described in the next section.

\section{GPU Porting} \label{sec:porting}

The popularity of GPUs in HPC has grown exponentially over the last decade, largely due to their ability to deliver massive performance. With the emergence of new vendors, developers have more freedom to choose the direction of their code development. This could mean choosing a particular GPU hardware and programming approach that is more suited to their numerical implementations. However, in the context of the latest pre-exascale/exascale supercomputers, which tend to use different GPU hardware, this approach is very limiting. Ideally, scientific codes should have an implementation that runs on any hardware, or in other words, be truly portable. This will allow GPU codes to approach and exceed the exascale limit on any architecture, and in turn solve computationally expensive problems that were previously thought to be infeasible. In this section we discuss the approach we have taken to exploit the features of STREAmS-2 to make it a truly GPU-portable solver. 

\subsection{Code architecture}

STREAmS was previously adapted to the changing HPC environment when it was ported to STREAmS-1 using CUDA Fortran to target NVIDIA GPUs, and a detailed description of the porting strategy and performance is available in the reference publication~\cite{bernardini2021streams}. STREAmS-1 was developed in Fortran primarily for its simplicity and improved code readability. In addition, there was minimal use of external libraries and heavy reliance on standardised frameworks (e.g., MPI). This allowed us to develop the code in a vendor-independent way, using multiple compilers and running on different supercomputers. However, the code remained largely procedural, preventing easy extension to other programming paradigms. In addition, STREAmS-1 made extensive use of \textit{ifdef} directives to separate the CPU and GPU (CUDA Fortran) versions of the code. This strategy is similar to other popular CFD codes \cite{zhu2018afid}\cite{costa2021gpu}, although STREAmS-1 uses a more distinct pattern to improve readability. Using the \textit{ifdef} approach forces developers to maintain both execution branches, which limits scaling to other architectures and seriously compromising code readability and maintainability. With the advent of newer GPU architectures, this development model needed to be adapted. 

STREAmS-1 was further developed into STREAmS-2 to provide a more scalable version, which further helped to solve the portability problem \cite{bernardini2023streams}. The design of STREAmS-2 is based on the object-oriented Fortran 2008 standard, which allows developers to accommodate completely different implementations depending on the backend, without having to duplicate or deeply restructure the entire code. Some of the ideas in this approach are similar to those used in Neko \cite{jansson2021neko}, although a more fine-grained object-oriented design is used in that case.

STREAmS-2 has been developed using CUDA Fortran to target NVIDIA GPUs, however, to extend the solver to other backends, five approaches were considered:

\begin{enumerate}
    \item Vendor specific paradigms: CUDA/HIP/OneAPI \cite{oneapi}
    \item OpenCL standard
    \item Directive-based standards: OpenMP \cite{openmp}/OpenACC
    \item Other high-level portability approaches: Kokkos \cite{kokkos}/Legion \cite{legion}/OpenSYCL \cite{opensycl}/alpaka \cite{alpaka}/RAJA \cite{raja}
    \item Intrinsic language constructs (\textit{do concurrent} from Fortran or \textit{parallel stl} from C++)
\end{enumerate}

Each of these strategies has its advantages and disadvantages. Option 3 is attractive because it is standardised, multi-device capable and high level, but the implementation is still lagging behind and advanced optimisations of complex kernels can be difficult. The same goes for 5, where the implementation is currently even more inadequate. Approach 2 is potentially good, but the resulting code is quite verbose and optimal performance is difficult to achieve. Option 4 is promising, but requires significant code rewrites, and support for different devices can be a critical issue.

In general, vendor-specific paradigms are still the best choice for achieving optimal performance. Since our specific goal is to enable the code to run on a large subset of modern HPC centres, the main targets are CUDA (NVIDIA) and HIP (AMD). The CUDA and HIP paradigms are essentially very similar, and the hipify \cite{hipify} tool makes it easy to convert code from CUDA to HIP. However, there are some critical issues with a Fortran based implementation. CUDA has a specific Fortran declination, CUDA Fortran, while HIP only offers HIPFort \cite{HIPFort}, which is basically a collection of Fortran variables and interfaces to HIP library functions. This means that at least two main strategies can be adopted:
\begin{enumerate}
    \item Develop using CUDA Fortran and periodically translate the code to the HIPFort+HIP paradigm.
    \item Develop a C/C++ implementation of the CUDA parts and write Fortran interfaces to call them. This means that hipify can be used to convert CUDA code to HIP.
\end{enumerate}
Strategy 1 makes CUDA Fortran development much easier, but HIP translation is difficult. Strategy 2 makes CUDA development harder, but HIP compilation easier. Given the normal development cycle of the code, which includes implementation of new methods and physics, we believe that an easier development approach is crucial. CUDA Fortran seems to be the best compromise between code readability and resulting performance. Therefore we decided to follow strategy 1. Since STREAmS-2 already uses an implementation of CUDA Fortran for the NVIDIA backend, this strategy essentially means porting the existing kernels directly to HIPFort+HIP to target the AMD backend. Note that the HIPFort backend can in principle also target the NVIDIA backend, but in this work we focus our development mainly on handling the AMD GPUs.

So far we have not mentioned the traditional CPU backend, which is generally important to support in a scientific code. We use a similar strategy for developing the CPU backend using the CUDA Fortran version, although the approach is much simpler due to the many similarities between them. 

\begin{figure}[]
    \centering
    \includegraphics[scale=0.5]{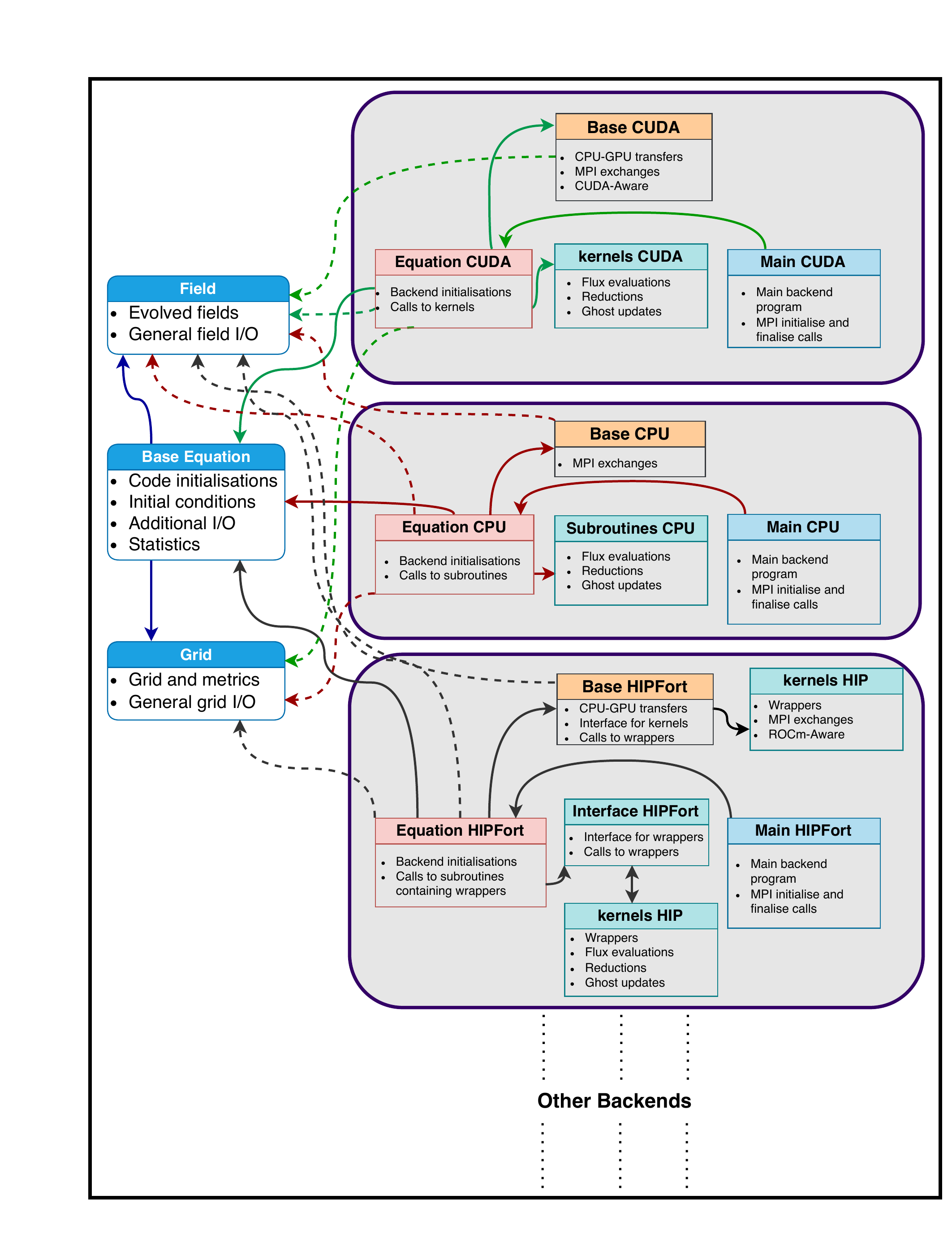}
    \caption{Program flow of STREAmS-2 with CUDA Fortran, HIPFort and CPU backends. This is an example of a single equation type with multiple program backends.  Blocks with labels represent Fortran objects or modules: Field and Grid (equation and backend independent), Base Equation (equation dependent and backend independent), Base CUDA/CPU/HIPFort (equation independent and backend dependent) and other blocks that are both equation and backend dependent. On the right, the grey container boxes group the blocks used by each backend. Each line represents an object that uses another object. Solid lines represent objects that contain other objects. Dashed lines represent objects that use other external objects but are saved internally as pointers.}
    \label{fig:streams_block}
\end{figure}
The final structure of the updated STREAmS-2 including support for three backends (CPU, CUDA Fortran and HIPFort) is shown in figure~\ref{fig:streams_block}. All the backends begin with a main program, which is essentially the entry point for running a particular backend and governing equation system. Further program structure is based on the following considerations,

\subsubsection{Equation and backend independent code}
This consists of the Field and Grid objects. The Grid object primarily uses procedures for general grid management, which mainly correspond to I/O routines and evaluation of grid metrics. The Field object, on the other hand, represents a generic field development of the solver along with I/O management. This part of the code also consists of a parameter module, which is mainly used to define precision, constants and other utility functions. 

\subsubsection{Equation Independent and backend dependent code}
This mainly concerns generic MPI communication procedures implemented according to a particular backend. The Base $\langle$backend$\rangle$ blocks in figure~\ref{fig:streams_block} represent this. Both the CUDA Fortran and HIPFort backends have some form of GPU-aware transfers, while this aspect is irrelevant in the CPU backend. 

\subsubsection{Equation dependent and backend independent code}
Next, the Base Equation block in figure~\ref{fig:streams_block} is dedicated to the initialisation of the flow and to perform all typical backend independent operations (e.g., statistics preparation, transport properties definition) for a particular equation type. This block can generally be extended to systems of other governing equations.

\subsubsection{Equation and backend dependent code}
The Equation $\langle$backend$\rangle$ block contains the main timeloop for a particular equation type. This block also varies for different backends. For example, the Equation CUDA block consists of all calls to kernels written in CUDA Fortran. For the CPU backend,  these kernels are the standard Fortran subroutines targeted at the CPU architectures. Finally, in the HIPFort backend, the Equation HIPFort further calls the wrappers required to execute the HIP kernels through C-Fortran interoperations.

\subsection{HIPFort implementation}
Our approach to the HIPFort implementation follows the structure used in the CUDA Fortran backend. This means that much of the CPU memory management, I/O routines and MPI implementation will remain largely the same and this is naturally achieved by exploiting the code separation guaranteed through the described object oriented approach. The majority of the changes for HIPFort come from the creation of the GPU memory allocations, the wrapper interface and finally the HIP kernels written in C/C++. The first fundamental change between CUDA Fortran and HIPFort is the creation of the GPU device arrays. For CUDA Fortran we create allocatable arrays with the device tag, but for HIPFort we create a pointer of a similar dimension. To better understand these differences, we provide an example of the creation and use of the \textit{w\_gpu} device array containing the conserved variables. 

Listing~\ref{listing:create} shows the main differences in creating device arrays using the CUDA Fortran and HIPFort approaches. Here \textit{rkind} corresponds to \textit{real64} or \textit{real32} Fortran 2008 predefined precision types that can be chosen before compilation. All benchmarks reported in this paper refer to the double precision \textit{real64} type.

\begin{chart}[H]
\begin{tcolorbox}
\textbf{CUDA Fortran}
\begin{lstlisting}[language=Fortran,columns=fullflexible]
real(rkind), allocatable, dimension(:,:,:,:), device :: w_gpu
\end{lstlisting}
\textbf{HIPFort}
\begin{lstlisting}[language=Fortran,columns=fullflexible]
real(rkind), pointer, dimension(:,:,:,:) :: w_gpu
\end{lstlisting}
\end{tcolorbox}
\caption{Creation of device arrays}
\label{listing:create}
\end{chart}

Listing~\ref{listing:alloc} illustrates the allocation methods. In CUDA Fortran we use the standard \textit{allocate} statement. For HIPFort we use the \textit{hipMalloc} routine. The \textit{hipcheck} routine checks the return status of the passed routines. This allocation of the main device array in the order $(x,y,z,v)$, where $x$, $y$, $z$ are the three spatial coordinates and $v$ the conserved variables, guarantees CUDA/HIP thread coalescence when thread parallelisation involves the first array dimension $x$. For the calculation of convective fluxes in the $x$ direction, however, the first array dimension cannot be parallelised, and additional field arrays of different order $(y,x,z,v)$ are allocated and filled at each iteration, transposing the normal field array. More details on this implementation can be found in \cite{bernardini2021streams}.

\begin{chart}[]
\centering
\begin{tcolorbox}
\textbf{CUDA Fortran}
\begin{lstlisting}[language=fortran,columns=fullflexible]
allocate(w_gpu(1-ng:nx+ng, 1-ng:ny+ng, 1-ng:nz+ng, nv))
\end{lstlisting}
\textbf{HIPFort}
\begin{lstlisting}[language=fortran,columns=fullflexible]
call hipCheck(hipMalloc(w_gpu,(nx+ng)-((1-ng))+1,(ny+ng)-((1-ng))+1,(nz+ng)-((1-ng))+1,nv))
\end{lstlisting}
\end{tcolorbox}
\caption{Allocation of device arrays}
\label{listing:alloc}
\end{chart}

A corresponding CPU array, \textit{w\_cpu}, is used during the code initialisation and finalisation procedures. Before time evolution is initiated, the CPU array is transferred to the GPU device memory. Listing~\ref{listing:transfer} illustrates the differences between the two approaches. 

\begin{chart}[]
\centering
\begin{tcolorbox}
\textbf{CUDA Fortran}
\begin{lstlisting}[language=fortran,columns=fullflexible]
w_gpu = w_cpu
\end{lstlisting}
\textbf{HIPFort}
\begin{lstlisting}[language=fortran,columns=fullflexible]
call hipCheck(hipMemcpy(w_gpu,w_cpu,hipMemcpyHostToDevice))
\end{lstlisting}
\end{tcolorbox}
\caption{CPU to GPU transfers}
\label{listing:transfer}
\end{chart}

For kernel implementation, CUDA Fortran employs two types of kernels. The most computationally intensive kernels are explicitly written with a block and grid configuration at call (global attribute kernels), while the relatively simple kernels are automatically generated using the kernel loop directive (\textit{cuf} kernels) feature of CUDA Fortran. For HIPFort, however, all kernels must be written explicitly as there is no native support for directive-based methods. In addition to the main kernel, a wrapper on the C/C++ side and an interface on the Fortran side are required. In listing~\ref{listing:updateflux1} we take an example of a simple flux update kernel to explain the main differences. The CUDA Fortran version consists of a simple \textit{cuf} kernel with directives placed before the loop and specifying the number of nested loops that are parallelised. 

For the HIPFort backend, we call a kernel wrapper and pass the corresponding C address object of the device arrays as an argument using Fortran's native \textit{c\_loc} procedure. The implementation of this kernel wrapper call is constituted by three components: an interface, a wrapper and finally the kernel. Listing~\ref{listing:updateflux2} provides the details of this implementation. The interface is  created to specify the arguments and the data type that must be compatible for Fortran-C interoperability. On the C/C++ side, we have a wrapper with arguments that match the interface on the Fortran side, and the grid and block definitions needed for the thread invocation. The vector (THREAD\_X,THREAD\_Y,THREAD\_Z) represents the block coordinates, which can be different for each kernel. This wrapper will eventually call the HIP kernel which will perform the required operations. Note that the first argument of the wrapper corresponds to the kernel stream. In this case, we use the default stream to launch the kernel, therefore, a \textit{c\_null\_pointer} is passed from the Fortran side to enforce this.

\begin{chart}[]
\centering
\begin{tcolorbox}
\textbf{CUDA Fortran}
\begin{lstlisting}[language=fortran,columns=fullflexible]
subroutine update_field_cuf(nx,ny,nz,ng,nv,w_gpu,fln_gpu)
  integer :: nx,ny,nz,nv,ng
  real(rkind), dimension(1-ng:,1-ng:,1-ng:,1:), intent(inout), device :: w_gpu
  real(rkind), dimension(1:,1:,1:,1:), intent(in),  device :: fln_gpu
  integer :: i,j,k,m  
  !$cuf kernel do(3) <<<*,*>>> 
  do k=1,nz
    do j=1,ny
      do i=1,nx
        do m=1,nv
          w_gpu(i,j,k,m) = w_gpu(i,j,k,m)+fln_gpu(i,j,k,m)
        enddo
      enddo
    enddo
  enddo
endsubroutine update_field_cuf
\end{lstlisting}
\textbf{HIPFort}
\begin{lstlisting}[language=fortran,columns=fullflexible]
subroutine update_field_kernel(nx,ny,nz,ng,nv,w_gpu,fln_gpu)
  integer ::  nx,ny,nz,nv,ng
  real(rkind), dimension(:,:,:,:), target :: w_gpu
  real(rkind), dimension(:,:,:,:), target :: fln_gpu
  call update_field_kernel_wrapper(c_null_ptr,nx,ny,nz,nv,ng,c_loc(w_gpu),c_loc(fln_gpu))
endsubroutine update_field_kernel
\end{lstlisting}
\end{tcolorbox}

\caption{A general comparison of kernel subroutines between CUDA Fortran and HIPFort. The CUDA Fortran implementation uses a \textit{cuf} directive for automatic kernel generation. The HIPFort implementation has a call to the wrapper which invokes a HIP kernel.}
\label{listing:updateflux1}
\end{chart}

\begin{chart}[]
\centering
\begin{tcolorbox}
\textit{(a) Interface}:
\begin{lstlisting}[language=fortran,columns=fullflexible]
interface
subroutine update_field_kernel_wrapper(stream,nx,ny,nz,nv,ng,w_gpu,fln_gpu)&
bind(c,name="update_field_kernel_wrapper")
  import :: c_ptr,c_int
  implicit none
  type(c_ptr), value :: stream
  integer(c_int), value :: nx,ny,nz,nv,ng
  type(c_ptr), value :: w_gpu,fln_gpu
end subroutine update_field_kernel_wrapper
end interface
\end{lstlisting}
\textit{(b) Wrapper}: 
\begin{lstlisting}[language=c,columns=fullflexible]
extern "C"{
void update_field_kernel_wrapper(hipStream_t stream,int nx,int ny,int nz,int nv,int ng,
                                 real *w_gpu,real *fln_gpu){
  dim3 block(THREAD_X,THREAD_Y,THREAD_Z);
  dim3 grid(divideAndRoundUp((nx)-(1)+1,block.x),
            divideAndRoundUp((ny)-(1)+1,block.y),
            divideAndRoundUp((nz)-(1)+1,block.z));

  hipLaunchKernelGGL((update_field_kernel),grid,block,0,stream,nx,ny,nz,nv,ng,
                                           fluid_mask_gpu,w_gpu,fln_gpu);
  }
}
\end{lstlisting}
\textit{(c) Kernel}: 
\begin{lstlisting}[language=c,columns=fullflexible]
__global__ void  update_field_kernel(int nx,int ny,int nz,int nv,int ng,
  real *w_gpu,real *fln_gpu){
  int i  = 1+(threadIdx.x + blockIdx.x * blockDim.x);
  int j  = 1+(threadIdx.y + blockIdx.y * blockDim.y);
  int k  = 1+(threadIdx.z + blockIdx.z * blockDim.z);
  if(i <= nx && j <= ny && k <=nz){
    for(int m=1; m<nv+1; m++){
      w_gpu[__I4_W(i,j,k,m)] = w_gpu[__I4_W(i,j,k,m)]+fln_gpu[__I4_FLN(i,j,k,m)];
    }
  }
}
\end{lstlisting}
\end{tcolorbox}

\caption{Specification of interface, wrapper and kernel for the update\_flux kernel based on the HIPFort backend. \textit{(a)} Interface contains definitions for all the arguments to be passed to the wrapper. \textit{(b)} Wrapper contains the block and grid definitions with a kernel launcher. \textit{(c)} Kernel with the parallel loop index definitions along with the main operations.}
\label{listing:updateflux2}
\end{chart}

Here, similar to the \textit{rkind} parameter on the Fortran side, the \textit{real} data type for the HIPFort kernel and wrapper can be a \textit{float} or a \textit{double}. Note that in the HIP kernel, the indices of the multi-dimensional device arrays must be flattened to a one-dimensional array to maintain compatibility between the Fortran and C/C++ indexing systems. An example
of flattening for the \textit{w\_gpu} device array is provided in listing~\ref{listing:linear}.

\begin{chart}[]
\centering
\begin{tcolorbox}
\begin{lstlisting}[language=c,columns=fullflexible]
#define __I4_W(i,j,k,m) (((i)-(1-ng))+((nx+ng)-(1-ng)+1)*((j)-(1-ng))+
                        ((nx+ng)-(1-ng)+1)*((ny+ng)-(1-ng)+1)*((k)-(1-ng))+
                        ((nx+ng)-(1-ng)+1)*((ny+ng)-(1-ng)+1)*((nz+ng)-(1-ng)+1)*((m)-(1)))
\end{lstlisting}
\end{tcolorbox}
\caption{Linearisation of the multi-dimensional array \textit{w\_gpu} to one dimension.}
\label{listing:linear}
\end{chart}
Reduction operations are common in any CFD solver, and the strategy for porting reduction kernels is illustrated in the listings~\ref{listing:reduction1} and \ref{listing:reduction2}. For CUDA Fortran, the implementation is relatively straightforward when using \textit{cuf} directives, as the variable to be reduced is specified in the directive.

For the HIPFort implementation, similar to listing~\ref{listing:updateflux1}, we specify a subroutine to call the kernel wrapper (see listing~\ref{listing:reduction1}). We employ the hipCUB \cite{hipcub} library, which is part of the ROCm software stack, to perform the required reductions. The reduction is performed in two steps. First, a kernel populates the device work array (\textit{redn\_3d\_gpu}). Second, this array is passed to a hipCUB wrapper function for reduction. All reduction operations in STREAmS-2 are performed on arrays of the same size, which allows us to reuse the same array for all reduction operations, thus limiting memory usage. Also, similar to the linearisation of the \textit{w\_gpu}, we use the \textit{\_\_I3\_REDN\_3D} macro for the \textit{redn\_3d\_gpu} array.

\begin{chart}[]
\centering
\begin{tcolorbox}
\textbf{CUDA Fortran}
\begin{lstlisting}[language=fortran,columns=fullflexible]
subroutine compute_residual_cuf(nx,ny,nz,ng,nv,fln_gpu,dt,residual_rhou)
  integer :: nx,ny,nz,ng,nv
  real(rkind), intent(out) :: residual_rhou
  real(rkind), intent(in) :: dt
  real(rkind), dimension(1:nx, 1:ny, 1:nz, nv), intent(in), device :: fln_gpu
  integer :: i,j,k  
  residual_rhou = 0.0_rkind
  !$cuf kernel do(2) <<<*,*>>> reduce(+:residual_rhou)
  do k=1,nz
    do j=1,ny
      do i=1,nx
        residual_rhou = residual_rhou + (fln_gpu(i,j,k,2)/dt)**2
      enddo
    enddo
  enddo
endsubroutine compute_residual_cuf
\end{lstlisting}
\textbf{HIPFort}
\begin{lstlisting}[language=fortran,columns=fullflexible]
subroutine compute_residual_kernel(nx,ny,nz,ng,nv,fln_gpu,dt,residual_rhou,redn_3d_gpu)
  integer ::  nx,ny,nz,ng,nv
  real(rkind) ::  dt
  real(rkind) ::  residual_rhou
  real(rkind), dimension(:,:,:,:), target :: fln_gpu
  real(rkind), dimension(:,:,:), target :: redn_3d_gpu
  residual_rhou = 0.0_rkind
  call compute_residual_kernel_wrapper(c_null_ptr,nx,ny,nz,ng,nv,dt, &
                                       c_loc(fln_gpu),residual_rhou,c_loc(redn_3d_gpu))
endsubroutine compute_residual_kernel
\end{lstlisting}
\end{tcolorbox}
\caption{A general comparison of reduction procedures between CUDA Fortran and HIPFort. The CUDA Fortran implementation uses a \textit{cuf} directive with the reduction variable specified in the directive.}
\label{listing:reduction1}
\end{chart}

\begin{chart}[H]
\centering
\begin{tcolorbox}
\textit{(a) Interface}:
\begin{lstlisting}[language=fortran,columns=fullflexible]
interface
subroutine compute_residual_kernel_wrapper(stream,nx,ny,nz,ng,nv,dt, &
                                           fln_gpu,residual_rhou,redn_3d_gpu)&
bind(c,name="compute_residual_kernel_wrapper")
  import :: c_ptr,c_int,c_rkind
  implicit none
  type(c_ptr), value :: stream
  integer(c_int), value :: nx,ny,nz,ng,nv
  real(c_rkind), value :: dt
  type(c_ptr), value :: fln_gpu
  real(c_rkind) :: residual_rhou
  type(c_ptr), value :: redn_3d_gpu
end subroutine compute_residual_kernel_wrapper
end interface
\end{lstlisting}
\textit{(b) Wrapper}: 
\begin{lstlisting}[language=c,columns=fullflexible]
extern "C"{
void compute_residual_kernel_wrapper(hipStream_t stream,int nx,int ny,int nz,int ng,int nv,
                                     real dt,real *fln_gpu,
                                     real *residual_rhou,real *redn_3d_gpu){
  dim3 block(THREAD_X,THREAD_Y,THREAD_Z);
  dim3 grid(divideAndRoundUp((nx)-(1)+1,block.x),
            divideAndRoundUp((ny)-(1)+1,block.y),
            divideAndRoundUp((nz)-(1)+1,block.z));
 
  hipLaunchKernelGGL((compute_residual_kernel),grid,block,0,stream,nx,ny,nz,ng,nv,dt,
                                               fln_gpu,redn_3d_gpu);
  // hipCUB reduction for sum
  reduce<real, reduce_op_add>(redn_3d_gpu, nz*ny*nx, residual_rhou);

}
\end{lstlisting}
\textit{(c) Kernel}: 
\begin{lstlisting}[language=c,columns=fullflexible]
__global__ void  compute_residual_kernel_residual_rhou(int nx,int ny,
int nz,int ng,int nv,real dt,real *fln_gpu,real *redn_3d_gpu){
  int i  = 1+(threadIdx.x + blockIdx.x * blockDim.x);
  int j  = 1+(threadIdx.y + blockIdx.y * blockDim.y);
  int k  = 1+(threadIdx.z + blockIdx.z * blockDim.z);
  if(i <= nx && j <= ny && k <=nz){
    redn_3d_gpu[__I3_REDN_3D(i,j,k)] = 0.0;
    redn_3d_gpu[__I3_REDN_3D(i,j,k)] = ((fln_gpu[__I4_FLN(i,j,k,2)]/dt))*
                                       ((fln_gpu[__I4_FLN(i,j,k,2)]/dt));
  }
}
\end{lstlisting}
\end{tcolorbox}
\caption{Specification of Interface, wrapper and kernel for the compute\_residual\_kernel based on the HIPFort backend. \textit{(a)} Interface contains definitions for all the arguments to be passed to the wrapper. \textit{(b)} Wrapper contains the block and grid definitions with a kernel launcher followed by the hipCUB reduction routine for summation. \textit{(c)} Kernel mainly populates the reduction array which will be reduced by hipCUB.}
\label{listing:reduction2}
\end{chart}

\subsection{PyconvertSTREAmS - A portability tool}

The previous section highlights that many of the porting operations from CUDA Fortran to HIP can be automated. Hence, the development of PyconvertSTREAmS came about naturally. This portability tool has been designed in-house targeting STREAmS-2 and tackles the problem of handling code duplication across different programming paradigms.

The PyconvertSTREAmS tool is capable of analysing the latest CUDA Fortran version of STREAmS-2 and generating code for other backends. The tool was originally developed for the CPU backend and then extended to generate interfaces, wrappers and kernels for the HIPFort backend. In general, the tool is able to generate ready-to-compile source code, however, manual adjustments may be required for optimal performance. The tool is also designed to extend STREAmS-2 to other backends such as OpenMP in the future.

The converter is written in Python3. It makes extensive use of regular expressions to match, substitute and further translate the CUDA Fortran kernels to the required backend. Conceptually, the converter extracts all the information from the CUDA Fortran backend, breaks it down into smaller pieces that are later translated and reassembled into a backend of choice. It should be emphasised that we are not interested in automatic code generation, but in establishing a development strategy where developers maintain only one backend and periodically synchronise it across different supported backends.

Translation from CUDA Fortran to HIPFort could potentially be done using a tool called GPUFORT \cite{gpufort}, which is an interesting project but is at the moment still a research tool. This means that the translated code is not complete, or at least usually requires significant manual adjustments. Furthermore, the resulting code is not very readable, which prevents further modifications. Although this may increase the programming effort, we expect it to have a better impact in the future as more GPU software paradigms emerge. For these reasons, our goal was to create a tool that:

\begin{itemize}
    \item Works under certain code policies that are always met for the STREAmS-2 development path
    \item Produces a fully functional HIPFort and CPU backend from the CUDA Fortran version of the code, although some manual input may be required at startup.
    \item Produces a perfectly readable code, so can be easily modified if required
    \item Provides a framework for further extension to other backends
\end{itemize}

\subsubsection*{\underline{CUDA Fortran to HIPFort}}

\begin{figure}[]
    \centering
    \includegraphics[scale=0.55]{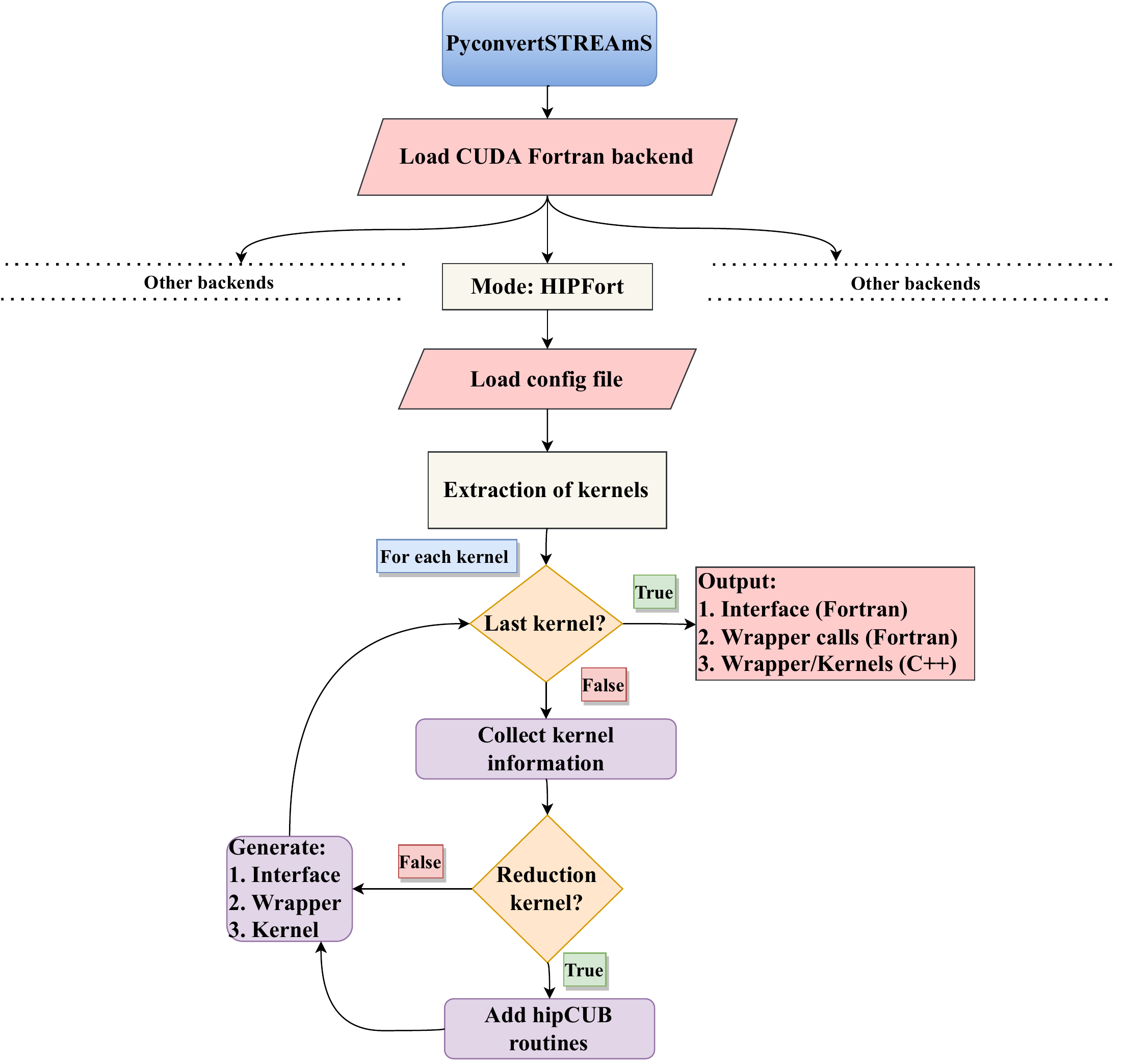}
    \caption{Program flow of PyConvertSTREAmS to obtain the HIPFort backend from the CUDA Fortran backend.}
    \label{fig:pyconvertstreams}
\end{figure}

Conversion from the CUDA Fortran backend to HIPFort is more complex than obtaining the CPU backend because the HIP kernels are not written in Fortran and therefore, in addition to the interface and wrappers supporting the kernel invocation, a Fortran to C/C++ translator has been developed. Figure~\ref{fig:pyconvertstreams} shows the program flow of the tool to obtain the HIPFort backend. Much of the development of this conversion was inspired by the design of GPUFORT (also the reduction implementation in listing~\ref{listing:reduction2}), although the translations to other backends are more automatic and generic. The tool also supports a config file capable of making changes or perform basic optimisations to specific kernels that are more suited to HIPFort framework. Some of them include:

\begin{itemize}
    \item Specifying the launch bounds 
    \item Changing the order of parallel loop indices
    \item Changing the extent of the loop parallelism
    \item Specifying the thread block configuration
\end{itemize}

Once the config file has been read, the tool proceeds to extract the device arrays, kernels and their attributes. It then loops through each of the kernels, translating to HIP and generating the corresponding interface and wrappers. In addition, if the kernel is of the reduction type, special hipCUB routines are added to the kernel wrapper. Finally, three files, Equation HIPFort, Interface HIPFort and Kernels HIP, corresponding to the HIPFort backend block in figure~\ref{fig:streams_block} are generated. This HIPFort backend compiles seamlessly and running the solver produces exactly the same results as the CUDA Fortran backend. 

\section{Performance analysis} \label{sec:results}

In this section we present the performance results of STREAmS-2 for the CUDA Fortran and HIPFort backends. The section is divided into two parts. In the first part, we evaluate the performance based on a single GPU card. The single-GPU performance analysis represents the fundamental evaluation of the overall code architecture -- in particular the memory layout -- and the computational kernels: different numerical conditions are evaluated against each other and ultimately against the peak performance of the hardware used. To achieve this, we use NVIDIA A100 and AMD MI250x GPUs, which form the computational core of several modern supercomputers. However, large production runs typically require codes to be run on multiple nodes with many GPUs. In the second part, we look at the scalability performance of the CUDA Fortran and HIPFort backends of STREAmS-2 based on two pre-exascale clusters that are part of the EuroHPC JU \cite{eurohpc}. 

\subsection{Single-GPU evaluation}

\begin{table}[H]
\centering
\begin{tabular}{ccccc}
\toprule
\multicolumn{1}{c}{Vendor} &
  \multicolumn{1}{c}{\begin{tabular}[c]{@{}c@{}}NVIDIA\end{tabular}} &
  \multicolumn{1}{c}{\begin{tabular}[c]{@{}c@{}}NVIDIA\end{tabular}} &
  \multicolumn{1}{c}{\begin{tabular}[c]{@{}c@{}}AMD\end{tabular}} &
  \multicolumn{1}{c}{\begin{tabular}[c]{@{}c@{}}AMD \end{tabular}} \\ 
  \toprule
  \multicolumn{1}{c}{Model} &
  \multicolumn{1}{c}{\begin{tabular}[c]{@{}c@{}} V100\\GPU\end{tabular}} &
  \multicolumn{1}{c}{\begin{tabular}[c]{@{}c@{}}A100\\GPU\end{tabular}} &
  \multicolumn{1}{c}{\begin{tabular}[c]{@{}c@{}}MI100\\GPU\end{tabular}} &
  \multicolumn{1}{c}{\begin{tabular}[c]{@{}c@{}}MI250x\\GCD\end{tabular}} \\ 
  \hline
Release year &
  2017 &
  2020 &
  2020 &
  2021 \\
  \hline
  Memory (GiB) &
  16 &
  40 &
  32 &
  64 \\
  \hline
\begin{tabular}[c]{@{}c@{}}Peak FP64 \\ performance\\  (TFLOPS)\end{tabular} &
  7.00 &
  9.70 &
  11.50 &
  23.90 \\
  \hline
\begin{tabular}[c]{@{}c@{}}Peak bandwidth\\  (GB/s)\end{tabular} &
  900.00 &
  1555.00 &
  1229.00 &
  1630.00 \\
  \hline
Approach &
  CUDA Fortran &
  CUDA Fortran &
  HIPFort &
  HIPFort \\
  \hline
  Compiler &
  \begin{tabular}[c]{@{}c@{}}HPC-SDK \\ 22.11\end{tabular} &
  \begin{tabular}[c]{@{}c@{}}HPC-SDK \\ 22.11\end{tabular} &
  \begin{tabular}[c]{@{}c@{}}hipfc/hipcc \\ ROCm 4.5.2\end{tabular} &
  \begin{tabular}[c]{@{}c@{}}hipfc/hipcc \\ ROCm 5.0.2\end{tabular}  \\
  \hline
Profiling &
  \begin{tabular}[c]{@{}c@{}}NVIDIA \\ Nsight Systems\end{tabular} &
  \begin{tabular}[c]{@{}c@{}}NVIDIA \\ Nsight Systems\end{tabular} &
  rocprof &
  rocprof \\
  \bottomrule
\end{tabular}
\caption{Description of the GPUs used in single-GPU performance evaluations.}
\label{tab:1}
\end{table}

\begin{figure}
 \begin{center}
  \includegraphics[scale=1,clip]{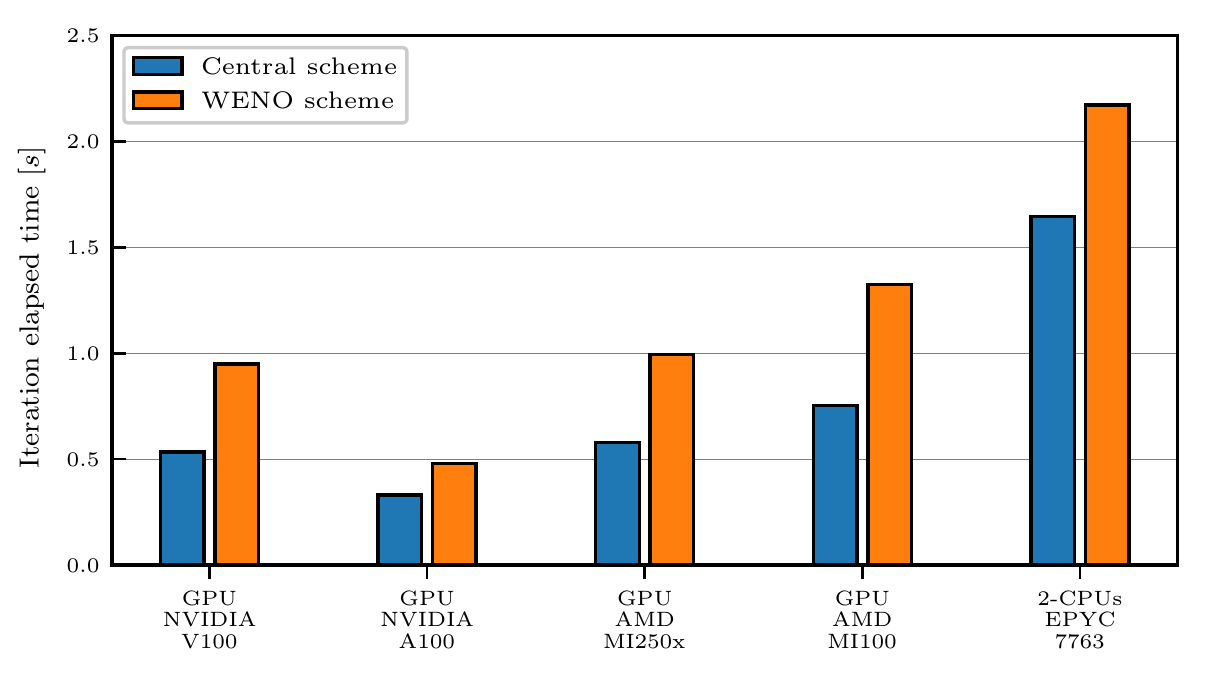}
  \caption{Elapsed time per iteration (s) for STREAmS-2 using different HPC architectures for grid: $(420 \times 250 \times 320)$. Comparisons between central and WENO flux evaluation schemes. CPU average time is based on a full node of 128 cores.}
  \label{fig:gpu_comp}
 \end{center}
\end{figure}

We evaluate the performance of STREAmS-2 on different CPU and GPU architectures (figure~\ref{fig:gpu_comp}). Table~\ref{tab:1} describes the configurations of the GPU architectures used for this plot. The grid size $(420 \times 250 \times 320)$ for each case was chosen with a maximum memory allocation on the V100 GPUs (16 GiB) that are part of the Marconi100 cluster. All simulations performed for figure~\ref{fig:gpu_comp} use a $6^{th}$ order central scheme and a $5^{th}$ order WENO scheme, which are generally adopted for production runs. Additional discussion of the impact of different orders of accuracy on run time will be provided later. For the CPU architecture, we consider a full compute node on the CPU partition of LUMI (LUMI-C) using MPI parallelisation. The decision to perform this study is not only to have comparable times between CPUs and GPUs, but also to ignore any intra-node CPU effects (which typically limit scalability) and to provide comparisons between independent units. For the GPUs, we consider a single card for NVIDIA V100, NVIDIA A100 and AMD MI100 GPUs. However, for AMD MI250x, we choose one of the two Graphics Compute Dies (GCDs) that make up a GPU card. The two GCDs are managed by two MPI processes in STREAmS-2 and the scheduler also identifies them as two GPU units.

GPUs are generally much faster than a single full CPU node on LUMI. In particular, GPU times range from two to four times faster than the CPU configuration under consideration. The trends in GPU results unsurprisingly follow their release dates (for the same vendor) and peak predictions. However, later in this section we will see that the real performance does not strictly follow the expected predictions as we try to understand the reasons for the observed discrepancies. For example, the performance of the MI250x GCD is similar to that of the old generation V100 GPU, although the peak performance is remarkably different. Comparing the results of the central and WENO schemes, we see that the GPU trends are very similar, while the CPU results for the central scheme are slightly worse when compared to the GPU data.

\begin{figure}
 \begin{center}
  \includegraphics[scale=1,clip]{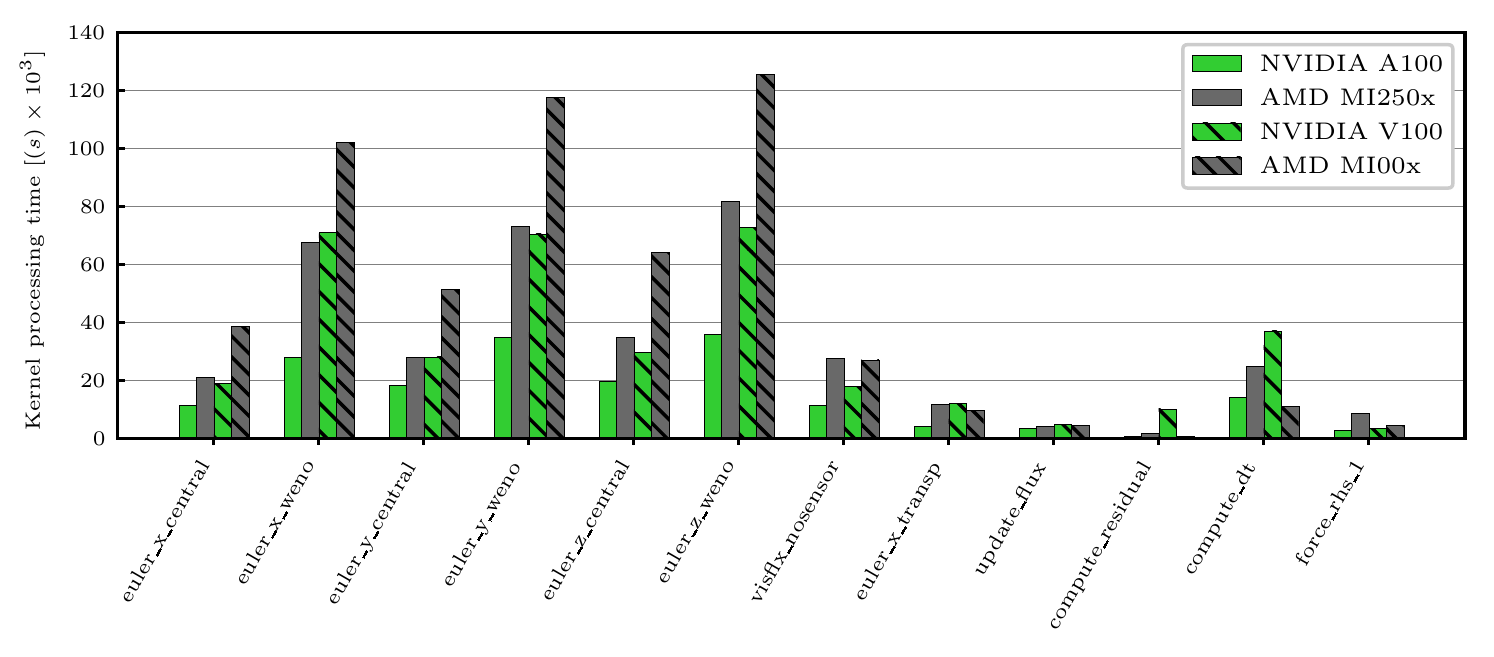}
  \caption{Kernel processing time ($s \times 10^3$) comparison for significant kernels of STREAmS-2 across four GPU architectures for grid: $(420 \times 250 \times 320)$. }
  \label{fig:kernel_comp}
 \end{center}
\end{figure}

\begin{table}[]
\centering
\begin{tabular}{cccc}
\toprule
\multirow{2}{*}{Kernel} & \multicolumn{2}{c}{Kernel type} & \multirow{2}{*}{Description} \\ \cline{2-3}
 & \multicolumn{1}{c}{CUDA Fortran} & HIPFort &  \\ \toprule
euler\_x\_central & \multicolumn{1}{c}{Global} & Global & \begin{tabular}[c]{@{}c@{}}Convective flux evaluation\\ in x-direction using\\ Central scheme\end{tabular} \\ \hline
euler\_x\_weno & \multicolumn{1}{c}{Global} & Global & \begin{tabular}[c]{@{}c@{}}Convective flux evaluation\\ in x-direction using\\ WENO scheme\end{tabular} \\ \hline
euler\_y\_central & \multicolumn{1}{c}{Global} & Global & \begin{tabular}[c]{@{}c@{}}Convective flux evaluation\\ in y-direction using\\ Central scheme\end{tabular} \\ \hline
euler\_y\_weno & \multicolumn{1}{c}{Global} & Global & \begin{tabular}[c]{@{}c@{}}Convective flux evaluation\\ in y-direction using\\ WENO scheme\end{tabular} \\ \hline
euler\_z\_central & \multicolumn{1}{c}{Global} & Global & \begin{tabular}[c]{@{}c@{}}Convective flux evaluation\\ in z-direction using\\ Central scheme\end{tabular} \\ \hline
euler\_z\_weno & \multicolumn{1}{c}{Global} & Global & \begin{tabular}[c]{@{}c@{}}Convective flux evaluation\\ in z-direction using\\ WENO scheme\end{tabular} \\ \hline
visflx\_nosensor & \multicolumn{1}{c}{cuf} & Global & \begin{tabular}[c]{@{}c@{}}Viscous flux\\ evaluation\end{tabular} \\ \hline
euler\_x\_transp & \multicolumn{1}{c}{cuf} & Global & \begin{tabular}[c]{@{}c@{}}Transposition of\\ main device array\end{tabular} \\ \hline
update\_flux & \multicolumn{1}{c}{cuf} & Global & Flux array update \\ \hline
compute\_residual & \multicolumn{1}{c}{cuf} & \begin{tabular}[c]{@{}c@{}}Global\\ hipCUB\end{tabular} & \begin{tabular}[c]{@{}c@{}}Residual evaluation\\ (sum reduction)\end{tabular} \\ \hline
compute\_dt & \multicolumn{1}{c}{cuf} & \begin{tabular}[c]{@{}c@{}}Global\\ hipCUB\end{tabular} & \begin{tabular}[c]{@{}c@{}}Time step evaluation\\ (max reduction)\end{tabular} \\ \hline
force\_rhs\_1 & \multicolumn{1}{c}{cuf} & \begin{tabular}[c]{@{}c@{}}Global\\ hipCUB\end{tabular} & \begin{tabular}[c]{@{}c@{}}Forcing term evaluations \\ (sum reduction)\end{tabular}  \\ 
\bottomrule
\end{tabular}
\caption{Description of the kernels used in figure~\ref{fig:kernel_comp}. CUDA Fortran employs global and cuf-directive kernels. HIPFort only employs global kernels with hipCUB implementation for reductions.}
\label{tab:kernel}
\end{table}

We now look at the individual kernel performance across the GPU architectures (figure~\ref{fig:kernel_comp}). Table \ref{tab:kernel} provides a description of the kernels used in the comparison. The results are reported in seconds unless otherwise stated in the figure. We have considered kernels based on three types that best represent the solver. The most computationally intensive kernels corresponding to convective and viscous flux evaluations (euler\_x\_central, euler\_x\_weno, euler\_y\_central, euler\_y\_weno, euler\_z\_central, euler\_z\_weno and visflx\_nosensor). The convective flux kernels represented by the central and WENO schemes are shown here as two different kernels (in each grid direction). In principle, however, the solver can be run in single (central and WENO) or hybrid mode. The effects of these modes are discussed in detail later in this section. These kernels are characterised by high complexity, especially for the WENO branch with high register usage, which can be crucial in limiting GPU occupancy and the final performance results. It should also be noted that the convective flux evaluations are explicitly implemented as global kernels for CUDA Fortran due to the use of local private arrays and device functions, but the viscous fluxes (visflx\_nosensor kernel) are implemented through \textit{cuf} directives in the CUDA Fortran backend. We also include some kernels that are not particularly demanding, but interesting for our implementation strategy and performance evaluations. These include a basic linear algebra kernel, update\_flux, and a transposition kernel, euler\_x\_transp. Finally, reduction kernels: compute\_residual (simple sum reduction), compute\_dt (max reduction) and force\_rhs\_1 (a more complex sum reduction) are used for the comparisons.

For convective flux kernels, performance is again comparable between the V100 GPU and the MI250x GCD, with the A100 GPU outperforming in all cases. The A100 GPU has an advantage of about 50\% for WENO executions and about 35\% for central executions. As the central execution is closer to basic linear algebra computation (compared to the more complex WENO code pattern), the results are expected to closely follow the peak bandwidth trends. Indeed, the update\_flux kernel, which is very similar to a simple matrix addition, shows close performance results for all devices with an average deviation of around 10\%, although the A100 GPU still outperforms the MI250x GCD despite the fact that the peak bandwidth should (minimally) favour the latter. The transposition kernel results are somewhat unexpected; while the V100 GPU and MI250x GCD results are almost identical, the MI100 GPU results are slightly better and, more interestingly, the A100 GPU results are about three times better. Optimal transposition kernels would require extensive use of shared memory and more advanced techniques. While \textit{cuf} directive kernels can automatically produce such optimised code, the reason for such a large difference between V100 and A100 GPUs is unclear. For the AMD counterparts, we decided to implement only naive transposition to maintain code clarity and simplify automation, as the impact of transposition on global run times is negligible. The results for reduction kernels (compute\_residual, etc.) are a little more difficult to analyse. While the superiority of the A100 GPU always holds, the results for other GPUs are more variable, but this is to be expected as different reduction implementations are involved. However, the only reduction kernel that is computationally relevant, compute\_dt, is typically called once every 10 iterations in production runs, so its contribution to the total run time is negligible. The last kernel of interest, visflx\_nosensor, although structurally simple (mainly derivatives and linear combinations of matrices), contains a fairly large number of lines (around 200). The performance advantage of NVIDIA is more evident in this case for all GPU generations. The results are analysed in more detail in the following sections.

\begin{table}[]
\centering
\begin{tabular}{ccccccccc}
\toprule
\multicolumn{1}{c}{\multirow{4}{*}{\begin{tabular}[c]{@{}c@{}}Launch \\ bounds\end{tabular}}} & \multicolumn{2}{c}{euler\_x} & \multicolumn{2}{c}{euler\_y} & \multicolumn{2}{c}{euler\_z} & \multicolumn{2}{c}{Iteration elapsed time} \\ \cline{2-9} 
\multicolumn{1}{c}{} & \multicolumn{1}{c}{\begin{tabular}[c]{@{}c@{}}A100\\GPU\end{tabular}} & \multicolumn{1}{c}{\begin{tabular}[c]{@{}c@{}}MI250x\\GCD\end{tabular}} & \multicolumn{1}{c}{\begin{tabular}[c]{@{}c@{}}A100\\GPU\end{tabular}} & \multicolumn{1}{c}{\begin{tabular}[c]{@{}c@{}}MI250x\\GCD\end{tabular}} & \multicolumn{1}{c}{\begin{tabular}[c]{@{}c@{}}A100\\GPU\end{tabular}} & \multicolumn{1}{c}{\begin{tabular}[c]{@{}c@{}}MI250x\\GCD\end{tabular}} & \multicolumn{1}{c}{\begin{tabular}[c]{@{}c@{}}A100\\GPU\end{tabular}} & \begin{tabular}[c]{@{}c@{}}MI250x\\GCD\end{tabular} \\ 
\cline{2-9}
& \multicolumn{8}{c}{Time(s) $\times 10^3$}  \\
\toprule
\multicolumn{9}{c}{central Scheme} \\ \toprule
\multicolumn{1}{c}{128} & \multicolumn{1}{c}{11.2} & \multicolumn{1}{c}{23.9} & \multicolumn{1}{c}{18.2} & \multicolumn{1}{c}{32.0} & \multicolumn{1}{c}{19.5} & \multicolumn{1}{c}{41.2} & \multicolumn{1}{c}{327.6} & 618.2 \\ \hline
\multicolumn{1}{c}{256} & \multicolumn{1}{c}{11.3} & \multicolumn{1}{c}{23.6} & \multicolumn{1}{c}{18.3} & \multicolumn{1}{c}{31.4} & \multicolumn{1}{c}{19.8} & \multicolumn{1}{c}{39.2} & \multicolumn{1}{c}{331.1} & 611.3 \\ \hline
\multicolumn{1}{c}{384} & \multicolumn{1}{c}{10.9} & \multicolumn{1}{c}{21.1} & \multicolumn{1}{c}{16.4} & \multicolumn{1}{c}{27.9} & \multicolumn{1}{c}{18.0} & \multicolumn{1}{c}{35.1} & \multicolumn{1}{c}{317.5} & 579.3 \\ \hline
\multicolumn{1}{c}{512} & \multicolumn{1}{c}{12.6} & \multicolumn{1}{c}{23.8} & \multicolumn{1}{c}{15.9} & \multicolumn{1}{c}{30.9} & \multicolumn{1}{c}{18.0} & \multicolumn{1}{c}{38.5} & \multicolumn{1}{c}{321.5} & 606.8 \\ \hline
\multicolumn{1}{c}{640} & \multicolumn{1}{c}{16.4} & \multicolumn{1}{c}{23.8} & \multicolumn{1}{c}{17.6} & \multicolumn{1}{c}{30.9} & \multicolumn{1}{c}{17.5} & \multicolumn{1}{c}{38.5} & \multicolumn{1}{c}{336.2} & 606.1 \\ \toprule
\multicolumn{9}{c}{WENO Scheme} \\\toprule
\multicolumn{1}{c}{128} & \multicolumn{1}{c}{27.8} & \multicolumn{1}{c}{71.8} & \multicolumn{1}{c}{34.7} & \multicolumn{1}{c}{71.4} & \multicolumn{1}{c}{35.8} & \multicolumn{1}{c}{75.6} & \multicolumn{1}{c}{476.8} & 984.2 \\ \hline
\multicolumn{1}{c}{256} & \multicolumn{1}{c}{27.9} & \multicolumn{1}{c}{72.0} & \multicolumn{1}{c}{34.9} & \multicolumn{1}{c}{70.4} & \multicolumn{1}{c}{35.8} & \multicolumn{1}{c}{75.3} & \multicolumn{1}{c}{477.8} & 982.7 \\ \hline
\multicolumn{1}{c}{384} & \multicolumn{1}{c}{28.3} & \multicolumn{1}{c}{67.7} & \multicolumn{1}{c}{32.2} & \multicolumn{1}{c}{73.1} & \multicolumn{1}{c}{37.7} & \multicolumn{1}{c}{81.4} & \multicolumn{1}{c}{476.4} & 994.5 \\ \hline
\multicolumn{1}{c}{512} & \multicolumn{1}{c}{36.8} & \multicolumn{1}{c}{73.4} & \multicolumn{1}{c}{39.7} & \multicolumn{1}{c}{71.0} & \multicolumn{1}{c}{43.5} & \multicolumn{1}{c}{81.0} & \multicolumn{1}{c}{542.7} & 100.43 \\ \hline
\multicolumn{1}{c}{640} & \multicolumn{1}{c}{50.5} & \multicolumn{1}{c}{122.3} & \multicolumn{1}{c}{48.0} & \multicolumn{1}{c}{92.2} & \multicolumn{1}{c}{46.2} & \multicolumn{1}{c}{87.6} & \multicolumn{1}{c}{616.0} & 1234.1 \\ \bottomrule
\end{tabular}
\caption{Effect of Launch bounds for grid: $(420 \times 250 \times 320)$. All launch bounds are taken as multiples of 128. Both the kernel processing time of euler\_x, euler\_y and euler\_z kernels and iteration elapsed time in $(s \times 10^3)$.}
\label{tab:launch_bounds}
\end{table}

To evaluate the behaviour of STREAmS-2 on different architectures, we perform some targeted tests. From here on, we only consider the A100 GPU and the MI250x GCD, which are more prevalent in HPC clusters at the time of writing. We start by measuring the impact of launch bounds on kernel run time (see table~\ref{tab:launch_bounds}). Convective kernels are characterised by high complexity and require a large number of registers. Under these conditions, fine-tuning the thread block can be effective in optimising execution. This is particularly true because launch bounds can be specified in the kernel definition itself to help the compiler decide on the number of registers that can be safely allocated. We consider the three most computationally demanding kernels and obtain the kernel and elapsed time for different thread block configurations and corresponding launch\_bounds declarations. We vary the thread configurations along $y$, while maintaining 128 threads for the $x$ block size. A separate test was performed with multiples of 64, but the performance was either similar or lower and is therefore not reported here. Note that since many parallelised loops span only two directions, we always fix the number of threads along $z$ as one.

For the central scheme, both the A100 GPU and the MI250x GCD have their lowest final run times for the $128 \times 3$ configuration, which is also reflected in the individual kernel run times. However, for the WENO scheme, the A100 GPU performs better with $128 \times 3$ and the MI250x GCD performs better with $128 \times 2$. For all further analysis in this section, we stick with the $128 \times 3$ configuration for two main reasons:
\begin{itemize}
    \item The difference between the configurations for the WENO scheme is less than 1\% as opposed to the central scheme where it is between 4-5\%
    \item Production runs of STREAmS-2 generally involve the use of hybrid schemes, and in most cases the central scheme forms the bulk of the kernel runs
\end{itemize}

\begin{table}[H]
\centering
\begin{tabular}{ccccc}
\toprule

 & \multicolumn{2}{c}{update\_flux} & \multicolumn{2}{c}{visflx\_nosensor} \\ \toprule
\multicolumn{1}{c}{\multirow{2}{*}{\begin{tabular}[c]{@{}c@{}}Parallel\\loops\end{tabular}}} & \multicolumn{1}{c}{\begin{tabular}[c]{@{}c@{}}A100\\GPU\end{tabular}} & \begin{tabular}[c]{@{}c@{}}MI250x\\ GCD\end{tabular} & \multicolumn{1}{c}{\begin{tabular}[c]{@{}c@{}}A100\\GPU\end{tabular}} & \begin{tabular}[c]{@{}c@{}}MI250x\\GCD\end{tabular} \\ 
\cline{2-5}
 & \multicolumn{4}{c}{Time(s) $\times 10^3$}  \\
 \cline{2-5}
\multicolumn{1}{c}{2} & \multicolumn{1}{c}{3.36} & 6.15 & \multicolumn{1}{c}{11.4} & 53.6 \\ \hline
\multicolumn{1}{c}{3} & \multicolumn{1}{c}{3.18} & 4.26 & \multicolumn{1}{c}{16.0} & 27.8 \\ \bottomrule
\end{tabular}
\caption{Effect of the number of parallel loops for grid: $(420 \times 250 \times 320)$. Kernel processing times for update\_flux and visflx\_nosensor kernels in $(s \times 10^3)$.}
\label{tab:parallel_loops}
\end{table}

In the next test we look at the effect of the number of parallel loops on the kernel run time (see table~\ref{tab:parallel_loops}). This drastically affects the number of CUDA/HIP threads invoked and therefore the number of operations performed by each thread. To illustrate this concept, we consider two kernels, update\_flux and visflx\_nosensor. As explained, the former is a minimal linear algebra kernel, while the latter is still structurally simple (no device function, no private local array) but has a much higher number of operations and consequently higher register usage.

For the CUDA Fortran backend, using \textit{cuf} kernel directives, the number of parallel loops has minimal impact on the update\_flux kernel. For visflx\_nosensor kernel, however, using only two parallel loops significantly improves the timing probably due to pipelining intra-kernel parallelisation and/or cache effects. On the other hand, for the HIPFort backend, parallelising all three loop indices leads to a dramatic improvement in performance, probably because the AMD compiler and devices are not able to take advantage of significant intra-kernel optimisations. Predicting the optimal number of parallelised loops in general is not straightforward and probably requires manual attempts.

\begin{table}[H]
\centering
\begin{tabular}{ccccccccc}
\toprule
\multicolumn{1}{c}{\multirow{4}{*}{Grids}} & \multicolumn{2}{c}{euler\_x} & \multicolumn{2}{c}{euler\_y} & \multicolumn{2}{c}{euler\_z} & \multicolumn{2}{c}{Grind time ($T_g$)} \\ \cline{2-9} 
\multicolumn{1}{c}{} & \multicolumn{1}{c}{\begin{tabular}[c]{@{}c@{}}A100\\GPU\end{tabular}} & \multicolumn{1}{c}{\begin{tabular}[c]{@{}c@{}}MI250x\\GCD\end{tabular}} & \multicolumn{1}{c}{\begin{tabular}[c]{@{}c@{}}A100\\GPU\end{tabular}} & \multicolumn{1}{c}{\begin{tabular}[c]{@{}c@{}}MI250x\\GCD\end{tabular}} & \multicolumn{1}{c}{\begin{tabular}[c]{@{}c@{}}A100\\GPU\end{tabular}} & \multicolumn{1}{c}{\begin{tabular}[c]{@{}c@{}}MI250x\\GCD\end{tabular}} & \multicolumn{1}{c}{\multirow{2}{*}{\begin{tabular}[c]{@{}c@{}}A100\\GPU\end{tabular}}} & {\multirow{2}{*}{\begin{tabular}[c]{@{}c@{}}MI250x\\GCD\end{tabular}}}
\\
\cline{2-7}
& \multicolumn{6}{c}{Time(s) $\times 10^3$}  \\
\toprule
\multicolumn{9}{c}{central Scheme} \\ \toprule
\multicolumn{1}{c}{Grid 1} & \multicolumn{1}{c}{26.1} & \multicolumn{1}{c}{13.3} & \multicolumn{1}{c}{18.7} & \multicolumn{1}{c}{30.6} & \multicolumn{1}{c}{19.5} & \multicolumn{1}{c}{38.4} & \multicolumn{1}{c}{0.91} & 1.72 \\ \hline
\multicolumn{1}{c}{Grid 2} & \multicolumn{1}{c}{25.8} & \multicolumn{1}{c}{13.0} & \multicolumn{1}{c}{15.9} & \multicolumn{1}{c}{29.8} & \multicolumn{1}{c}{17.8} & \multicolumn{1}{c}{33.8} & \multicolumn{1}{c}{0.91} & 1.73 \\ \hline
\multicolumn{1}{c}{Grid 3} & \multicolumn{1}{c}{28.2} & \multicolumn{1}{c}{14.4} & \multicolumn{1}{c}{20.8} & \multicolumn{1}{c}{33.2} & \multicolumn{1}{c}{21.9} & \multicolumn{1}{c}{39.8} & \multicolumn{1}{c}{1.09} & 1.96 \\ \hline
\multicolumn{1}{c}{Grid 4} & \multicolumn{1}{c}{21.1} & \multicolumn{1}{c}{11.2} & \multicolumn{1}{c}{18.3} & \multicolumn{1}{c}{27.9} & \multicolumn{1}{c}{19.8} & \multicolumn{1}{c}{35.1} & \multicolumn{1}{c}{0.98} & 1.72 \\ \toprule
\multicolumn{9}{c}{WENO scheme} \\ \toprule
\multicolumn{1}{c}{Grid 1} & \multicolumn{1}{c}{66.1} & \multicolumn{1}{c}{31.0} & \multicolumn{1}{c}{36.7} & \multicolumn{1}{c}{73.4} & \multicolumn{1}{c}{37.2} & \multicolumn{1}{c}{78.0} & \multicolumn{1}{c}{1.34} & 2.69 \\ \hline
\multicolumn{1}{c}{Grid 2} & \multicolumn{1}{c}{61.2} & \multicolumn{1}{c}{30.3} & \multicolumn{1}{c}{31.2} & \multicolumn{1}{c}{67.4} & \multicolumn{1}{c}{33.8} & \multicolumn{1}{c}{70.7} & \multicolumn{1}{c}{1.33} & 2.69 \\ \hline
\multicolumn{1}{c}{Grid 3} & \multicolumn{1}{c}{66.9} & \multicolumn{1}{c}{33.6} & \multicolumn{1}{c}{40.0} & \multicolumn{1}{c}{86.4} & \multicolumn{1}{c}{40.7} & \multicolumn{1}{c}{92.3} & \multicolumn{1}{c}{1.59} & 3.21 \\ \hline
\multicolumn{1}{c}{Grid 4} & \multicolumn{1}{c}{67.7} & \multicolumn{1}{c}{27.5} & \multicolumn{1}{c}{34.9} & \multicolumn{1}{c}{73.1} & \multicolumn{1}{c}{36.0} & \multicolumn{1}{c}{81.3} & \multicolumn{1}{c}{1.42} & 2.96 \\ \bottomrule
\end{tabular}
\caption{Effect of thread masking. Individual kernel times for euler\_x, euler\_y and euler\_z kernels are in $(s \times 10^3)$.}
\label{tab:grid}
\end{table}

Next, we test the effect of thread masking on the kernel run times of the three convective kernels (see table~\ref{tab:grid}). It is known that if the number of grid points is not a multiple of the thread block, we will have inactive threads in the warp (in the case of NVIDIA) or wavefront (in the case of AMD). Depending on the extent of the inactive threads, we can expect a reduction in performance. We measure the grind time as the comparisons are on different, albeit slightly different grids. The grind time ($T_g$), also known as the data processing rate, is the ratio of the run time and the number of GPUs ($N_{GPU}$) to the number of grid points ($N$) and is expected to be constant under ideal conditions. Mathematically, the grind time can be obtained as $T_g = T \times N_{GPU} / N$. Intuitively, this can be thought of as the time it takes one GPU to process one grid point.  In the table~\ref{tab:grid}, the grind time is given as the time to process one hundred million points. In the context of single GPU evaluations, $N_{GPU} = 1$, therefore we calculate the grind time as $T_g = T \times 10^8 / N$. In this test, we always use the optimal thread configuration of $128 \times 3$ (see table~\ref{tab:launch_bounds}). To test thread masking, we start by considering grids that are multiples of 128 (the largest CUDA/HIP thread size) and gradually modify these numbers to reproduce less ideal conditions:

\begin{itemize}
\item Grid 1: $(384 \times 256 \times 384)$ - Multiples of 128 in all three directions
\item Grid 2: $(384 \times 256 \times 350)$ - Multiples of 128 in two directions ($x$ and $y$)
\item Grid 3: $(420 \times 256 \times 350)$ - Multiples of 128 in one direction ($y$)
\item Grid 4: $(420 \times 250 \times 320)$ - Not a multiple of 128 in any direction
\end{itemize}

From the table~\ref{tab:grid}, as expected, grids 1 and 2 give the best times in most cases. For the A100 GPU, the reduction in performance between best and worst case is around 20\% for both the central and WENO schemes. For the MI250x GCD, the reduction is around 14\% for the central scheme and 20\% for the WENO scheme. Unexpectedly, the performance of grid 4 is generally better than grid 3. Overall, the use of a good thread block configuration is recommended for production runs where possible. 

\begin{figure}[H]
 \begin{center}
  \includegraphics[scale=1,clip]{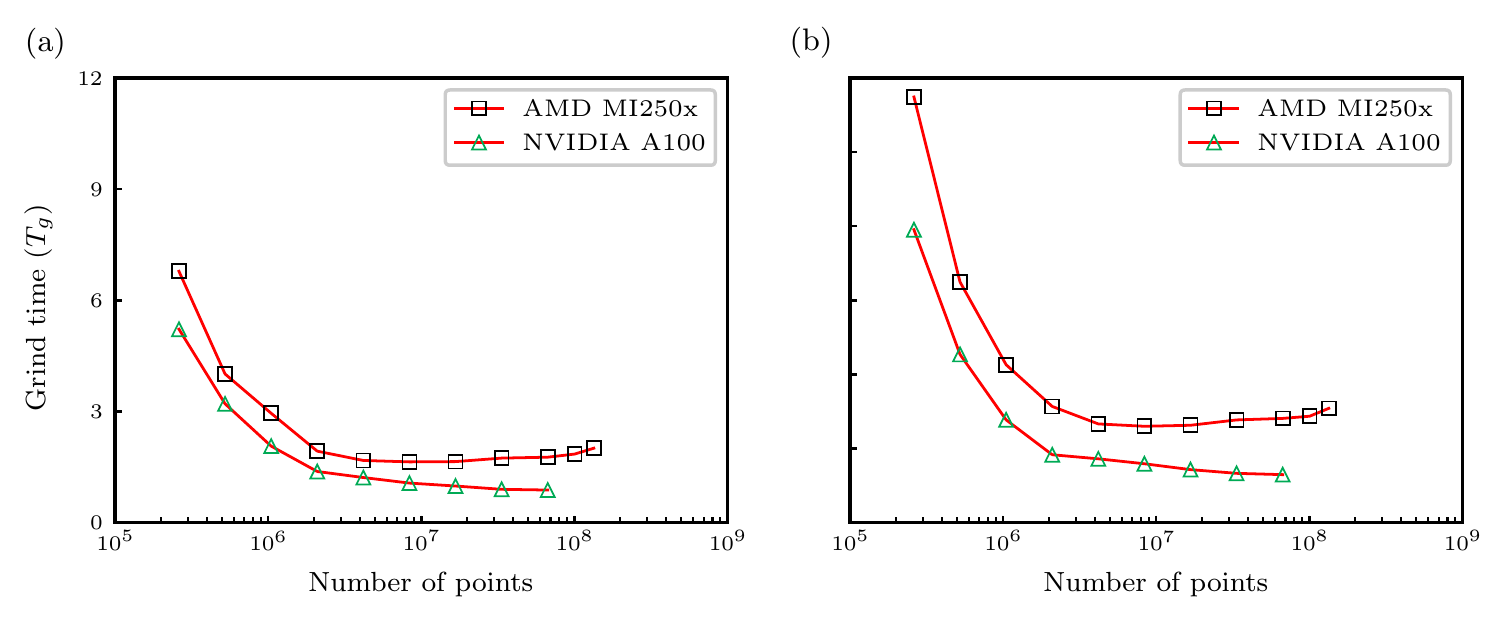}
  \caption{ Effect of grid size. \textit{(a)} central scheme. \textit{(b)} WENO scheme.}
  \label{fig:grind_time}
 \end{center}
\end{figure}

It is well known that GPUs perform well when processing large numbers of grid points. This is due to their inherent data parallelism. It is therefore interesting to measure performance in terms of the number of grid points, as this is a hard limit to strong scaling (where each GPU processes a progressively smaller part of the grid). Of course, since we are comparing different grids, we need to look at grind times ($T_g$) rather than total execution times. We use a base grid of around 0.25 million points and gradually double the grid size, with the largest grid being around 134 million points for the MI250x GCD and 67 million points for the A100 GPU. 

Figures~\ref{fig:grind_time}a and~\ref{fig:grind_time}b illustrate the effect of grid size for the central and WENO schemes respectively. The NVIDIA cards show a monotonous behaviour as the number of points increases. We can see two different ranges if we consider the logarithmic scale. In the first range, from 0.25 to 1 million points for the central scheme and from 0.25 to 2 million points for the WENO scheme, the reduction in grind time is enormous, proving that this range of points is not enough to exploit the parallelisation potential of the GPU. From 2 million grid points, there is still an improvement in grind time, but at a much lower rate. These two ranges are clearly visible for both central and WENO activation. All in all, the GPU delivers good results when the number of points to be processed exceeds 2 million. On the other hand, AMD's results show two trends: the first (number of points less than 2 million) is characterised by a sharp reduction in grind time, as is the case with its NVIDIA counterpart. The second, on the other hand, shows a slight increase in grind time. It turns out that there is an optimal number of grid points to obtain the best results on AMD cards, and it is not the largest possible case. The presented analysis also shows that the grid size chosen for most of the tests in this section is in the optimal range for the GPUs in question. 

\begin{figure}[H]
\begin{center}
\includegraphics[]{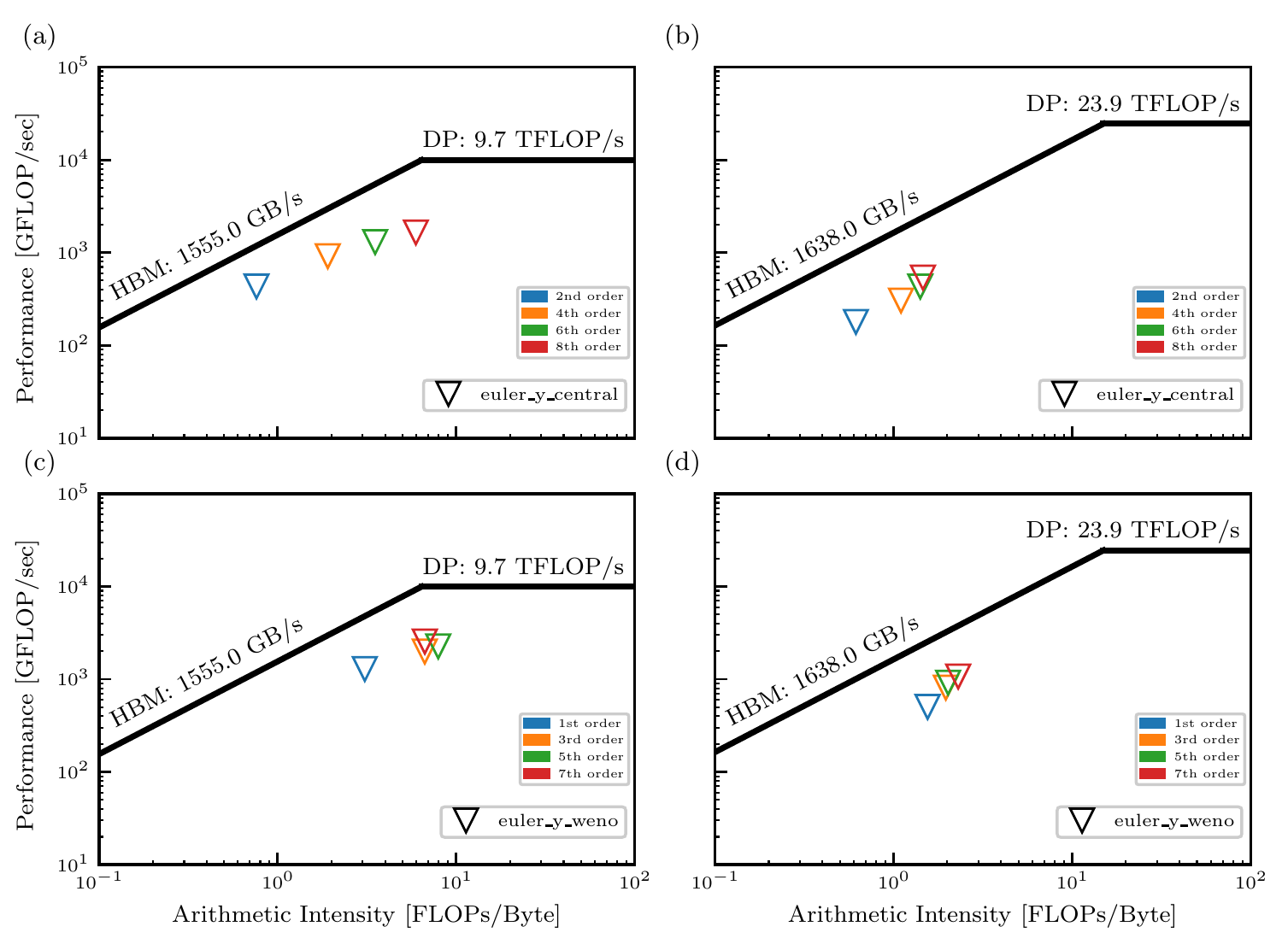}
\end{center}
\caption{Roofline analysis for euler\_y\_weno and euler\_y\_central kernels for grid: $(420 \times 250 \times 320)$. \textit{(a)} and \textit{(c)} AMD MI250x GCD. \textit{(b)} and \textit{(c)} NVIDIA A100 GPU. The marker corresponds to the kernel of the convective scheme, coloured according to its order of accuracy.}
 \label{fig:roofline}
\end{figure}

We performed a roofline analysis based on~\cite{yang2020hierarchical}. Roofline analysis is a very useful method to understand how the achieved performance compares to the peak performance of a device. It can also be used to identify bottlenecks, which can help with further optimisations. To obtain a roofline plot, the performance of a kernel is plotted against the Arithmetic Intensity (AI). AI is the ratio of floating point operations performed to data movement (FLOPs/Byte). Intuitively, a higher AI could represent better data locality. A roofline plot consists of two ceilings and a ridge. The region below the first ceiling, at lower AI, is limited by bandwidth, also known as the memory bounded region. Different memory layers could be considered, assuming data is found on High Bandwidth Memory (HBM), L2 cache and L1 cache, resulting in different memory ceilings. We will consider a standard roofline analysis based on HBM. The region below the second ceiling is limited by the peak performance of the device, also known as the compute bound region. Peak performance generally assumes that all operations are performed as Fused Multiply-Add (FMA), which is practically impossible to achieve for a real code. The ridge point is the transition point between the two ceilings.

Figures~\ref{fig:roofline}a-\ref{fig:roofline}d illustrate the roofline plots for the euler\_y kernel. Figures~\ref{fig:roofline}a and \ref{fig:roofline}c show the analysis for the central and WENO schemes based on the A100 GPU, respectively, while figures~\ref{fig:roofline}b and \ref{fig:roofline}d show the same for the MI250x GCD. The plots also include comparisons between four different orders of accuracy for the central ($2^{nd},4^{th},6^{th},8^{th}$) and WENO ($1^{st},3^{rd},5^{th},7^{th}$) schemes. 

\begin{figure}[H]
 \begin{center}
  \includegraphics[scale=1.0,clip]{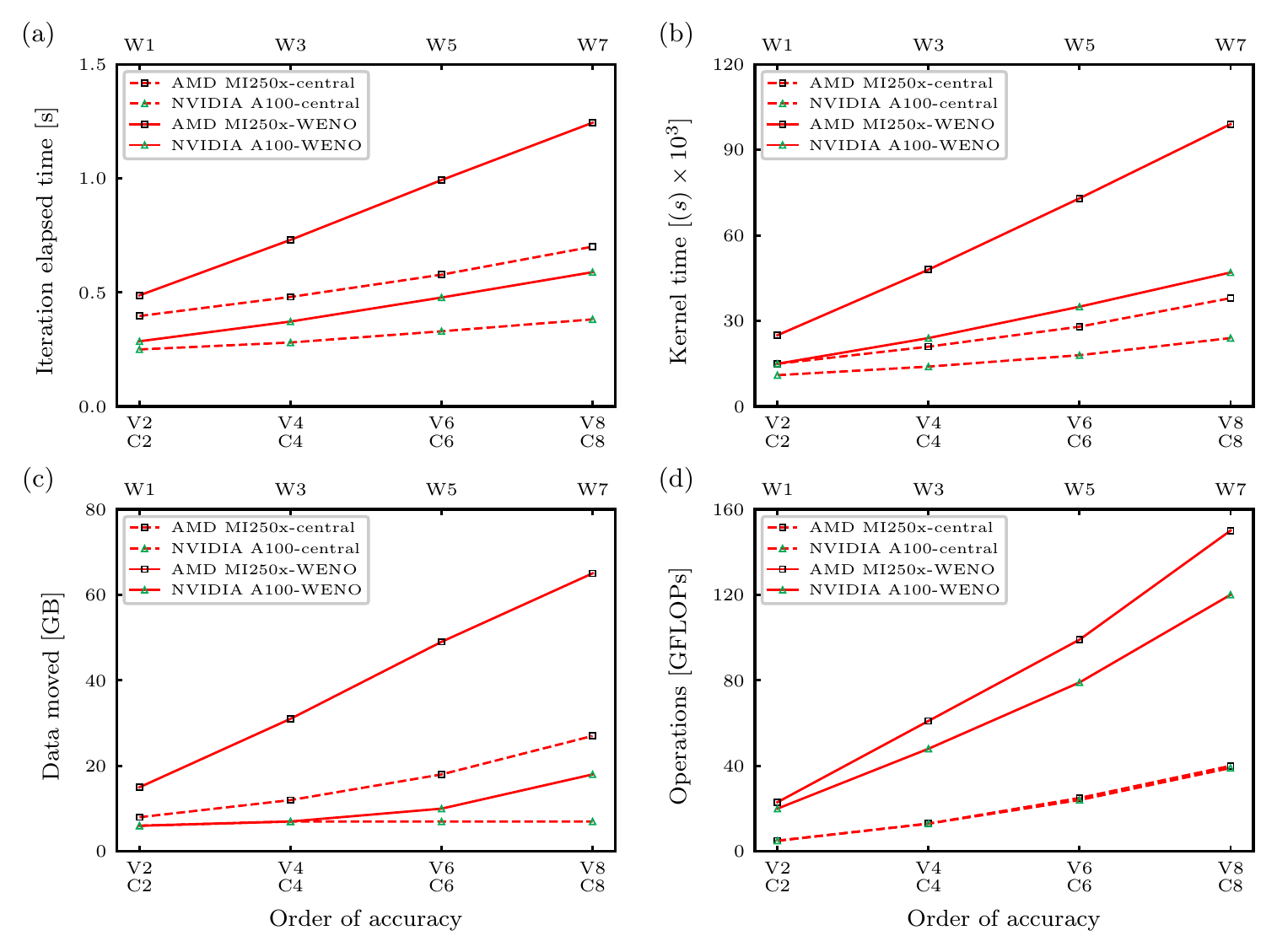}
  \caption{Effect of order of accuracy for grid: $(420 \times 250 \times 320)$. \textit{(a)} Effect on average time per iteration (s). \textit{(b)} Effect on euler\_y processing time $(s \times 10^3)$. \textit{(c)} Effect on the data moved to and from the HBM (GB) for the euler\_y kernel. \textit{(d)} Effect on the number of operations (GFLOPs) for the euler\_y kernel. All values taken for MI250x GCD and A100 GPU for central and WENO flux computation schemes. x-labels refer to the numerical accuracies of viscous terms ($V2,V4,V6,V8$) or convective central schemes ($C2,C4,C6,C8$) and convective WENO schemes ($W1,W3,W5,W7$). Dashed green lines correspond to central run time and solid green lines correspond to WENO run time for A100 GPU. Similarly, the black lines correspond to the MI250x GCD.}
  \label{fig:roofline2}
 \end{center}
\end{figure}

As expected for an explicit CFD code, we are generally in the memory bound regions. However, there are important differences. The MI250x GCD is always memory bound for both kernels and all orders of accuracy, whereas the A100 GPU approaches the computational bound regions at high orders of accuracy. For the A100 GPU, the increase in computational intensity becomes more pronounced as the order of accuracy increases. This is particularly true for the central schemes. WENO schemes, on the other hand, show relatively high AI for the A100 GPU at every order. The MI250x GCD, on the other hand, is always limited to an AI range of 1 to 3 and the corresponding performance is very limited. This may be related to the reduced data reuse (this is explained in more detail in figure~\ref{fig:roofline2}). As expected for realistic scientific codes, all points follow the behaviour of the roofline boundary, but are a little far from the ceiling. This means that there is potential room to approach the hardware limits. However, it should be noted that the floating-point limits take into account full FMA operations, which are unrealistically achievable for STREAmS-2. In terms of HBM measurements, caching generally dramatically improves the perceived memory bandwidth. In other words, from the programmer's point of view (memory accesses measured against source code), the bandwidth is significantly greater than the measured HBM bandwidth. In this context, possible future work could include additional memory optimisations (e.g., shared memory). However, it should be noted that STREAmS-2, as a community code in perpetual evolution, has to ensure reasonable readability and maintainability, even considering the possible addition of new numerics. Therefore, it is not always practical to develop a solver that is too machine specific or has complex optimisations.

To better understand the effect of the order of accuracy, we now look separately at elapsed times, total data movement to and from HBM and the number of operations performed. From a developer's point of view, these quantities are usually measured per second, as this is more meaningful for analysing performance. In this case, however, the quantities considered can be used to determine the capabilities of both the compilers and the caching mechanisms of the devices, which makes them interesting to study. The results are given for the total iteration time and for the euler\_y kernel at different numerical accuracies, considering both central and WENO modes, and for both GPUs. The bottom x-axis represents the order of accuracy for the discretisation of the convective central scheme (prefix C) and the viscous terms (prefix V), and the top x-axis represents the order of accuracy for the WENO reconstruction (prefix W). For example, W1 and C2 correspond to a first-order (i.e. upwind) WENO reconstruction and a second-order central scheme, respectively. Figure~\ref{fig:roofline2}a shows the iteration time of the solver with respect to the order of accuracy, while Figure~\ref{fig:roofline2}b shows the same for the euler\_y kernel. Figures~\ref{fig:roofline2}c and \ref{fig:roofline2}d show the movement of the HBM data and the total operations performed by the kernel as a function of the order of accuracy.

Figure~\ref{fig:roofline2}a shows a substantial linear behavior for both GPUs and both central/WENO activation. Time ratios among different codes are similar but show a small advantage for A100 GPU runs when increasing the accuracy order. For instance, at first order WENO, both AMD and NVIDIA times are much closer, but this progressively increases with the order of accuracy. To simplify the analysis, we look at the run times for a single euler\_y kernel. In figure~\ref{fig:roofline2}b we see an increase in kernel time for the euler\_y kernel that is virtually linear with increasing order of accuracy.  As already discussed, for the higher orders of accuracy, the absolute values are always significantly different comparing the A100 GPU and the MI250x GCD, with a large advantage for the NVIDIA card. The other plots in this figure allow us to understand the possible reasons for this difference, despite the fact that AMD has higher peak performance values. In figure~\ref{fig:roofline2}c we see that the data movement for the MI250x GCD is dramatically higher than for the A100 GPU in both cases. This is the main reason why NVIDIA shows a larger computational intensity as shown in the roofline analysis. As we only measure global memory accesses, this means that the NVIDIA GPU can extract the required data from registers and cache more often than its AMD counterpart. This is particularly true for the central scheme, which requires the same amount of data movement for each numerical order using the A100 GPU, while the MI250x GCD shows more than linear behaviour. Considering WENO, the A100 GPU still moves much less data, but the increase with numerical order is now more than linear for the A100 GPU, while a relatively perfect linear behaviour is visible for the MI250x GCD. In general, the different sizes of data moved benefit the A100 GPU's execution time greatly, as the peak bandwidth times for the two devices are almost similar. This is typically due to better compiler translation and a more efficient device caching mechanism. Figure~\ref{fig:roofline2}d shows the number of operations. The differences between AMD and NVIDIA are much less pronounced: there is almost no difference for the central scheme, while there is a moderate but measurable reduction in the number of operations for the A100 GPU when dealing with WENO schemes. All in all, as discussed for the roofline analysis, the code is mostly memory-bound and the main performance advantage of NVIDIA seems to be the large reduction in the amount of data moved, while from a programming point of view it consists of the same set of operations.

\begin{figure}[H]
 \begin{center}
  \includegraphics[scale=1.0,clip]{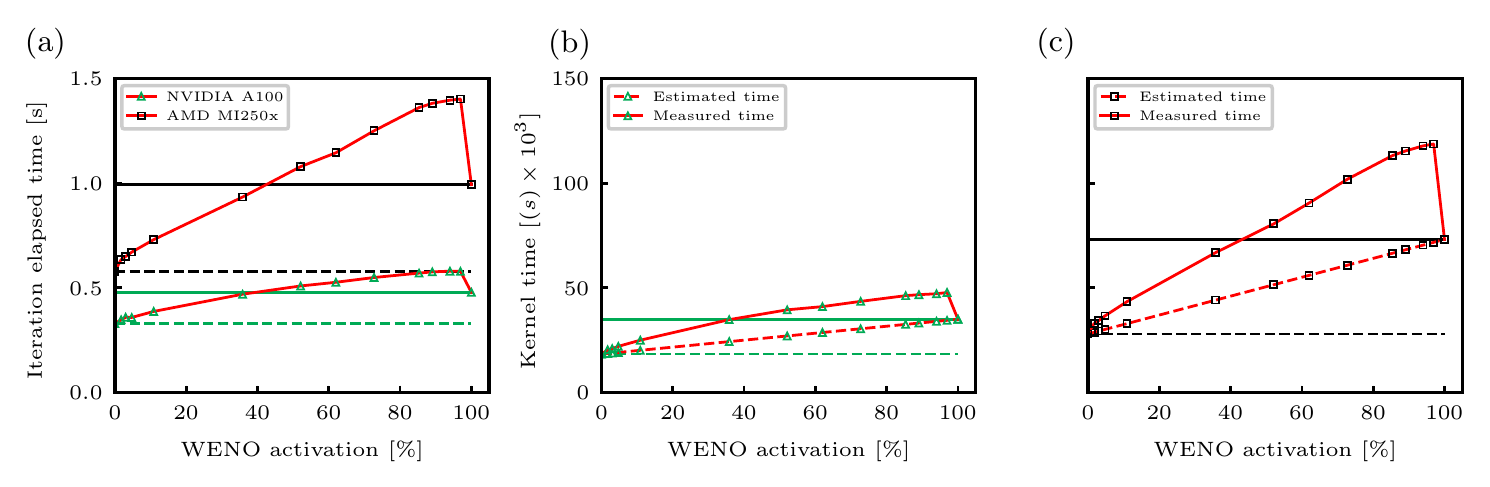}
  \caption{Effect of WENO activation on thread divergence for grid: $(420 \times 250 \times 320)$. \textit{(a)} Effect on average time per iteration (s). \textit{(b)} Effect on euler\_y kernel processing time for NVIDIA A100 GPU $(s \times 10^3)$. \textit{(c)} Effect on euler\_y kernel processing time for AMD MI250x GCD $(s \times 10^3)$.}
  \label{fig:hybrid}
 \end{center}
\end{figure}

STREAmS-2 production runs often include hybrid schemes (a mixture of central and WENO schemes). The percentage activation of the WENO kernel, which is a computationally demanding kernel, depends on the shock sensor threshold and the flow type. Therefore, the percentage of WENO activation is not known a priori and may even vary during computation. In order to understand its impact on the run times, we look at the performance of a kernel (euler\_y) and the full solver for different WENO activation percentages using the hybrid scheme.  

Figure~\ref{fig:hybrid}a shows the average iteration time as a function of the WENO activation percentage. The average iteration time increases as the WENO activation percentage increases. This is due to both the increase in operations due to the WENO scheme and the thread divergence. To see the effect of thread divergence, horizontal lines corresponding to central (dashed lines) and WENO (solid lines) execution times are also plotted for each GPU. As expected, the results are close to the central scheme for minimal WENO activations. However, it can be seen that as the activation approaches about 40\%, which corresponds to the full WENO execution time, the run times become larger as the activation increases. The highest observed time is about 20\% more than the full WENO time for A100 GPU, while it is about 40\% for MI250x GCD. At 100\% activation, full WENO timings are clearly recovered and this happens very abruptly.

Figures~\ref{fig:hybrid}b and \ref{fig:hybrid}c look separately at the effects of thread divergence for the A100 GPU and MI250x GCD respectively using the euler\_y kernel, where the hybrid schemes have a huge impact. The horizontal lines here, similar to \ref{fig:hybrid}a, correspond to the central (dashed lines) and WENO (solid lines) execution times, but for the kernel in question. In addition to the measured time (solid lines), we also plot the estimated time (dashed lines). The estimated time is calculated as the linear combination of the central and WENO run times, weighted by the WENO activation value. It turns out that the equilibrium point (hybrid run time equal to full WENO run time) is around 40\% of the activation for both GPUs. Also, the thread divergence effect is up to 30\% of the kernel time. All in all, both the NVIDIA and AMD GPUs show an effect of thread divergence but the extent seen is lower in the NVIDIA GPUs. 

\subsection{Scalability}
The performance of an efficient DNS code ultimately depends on how well it scales across a multi-node architecture, where each node contains multiple GPUs. We evaluate the inter-node parallel performance of STREAmS-2 based on the CUDA Fortran and HIPFort backends. Scalability is measured on LUMI (CSC) equipped with AMD MI250x GPUs and Leonardo (Cineca) equipped with NVIDIA A200 GPUs (Leonardo A200 is a slightly enhanced version of the A100 used for the previous single-GPU analysis). 

A common bottleneck in multi-node simulations is the MPI communication required to evaluate the derivatives. In this work, we use a GPU-aware communication model to perform these exchanges efficiently, avoiding unnecessary transfers between CPU and GPU. STREAmS-2 can perform this communication in two modes, synchronous and asynchronous, and more details on these two communication patterns are available in the reference publication of STREAmS-1~\cite{bernardini2021streams}.  We evaluate the performance differences between these two modes. In addition, we present the differences between the central and WENO flux evaluation modes.

\begin{figure}[H]
    \centering
    \includegraphics{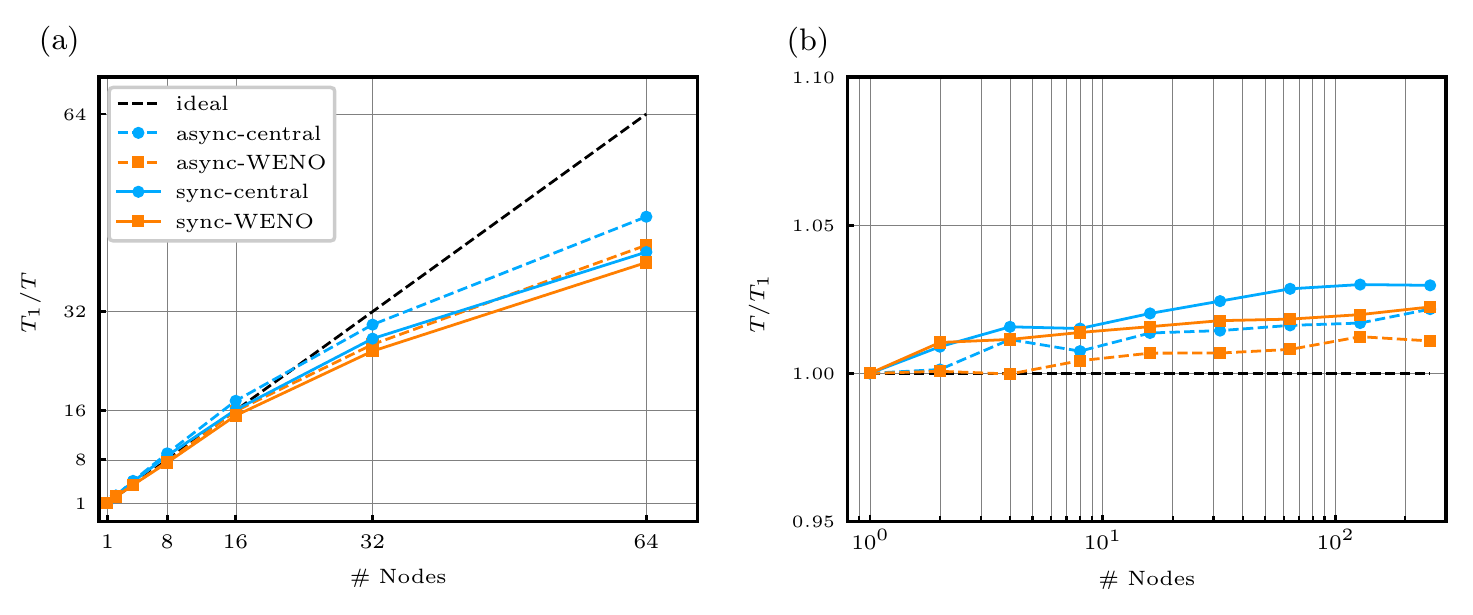}
    \caption{Strong \textit{(a)} and weak \textit{(b)} scalability plots for STREAmS-2 in synchronous and asynchronous modes for central and WENO flux schemes on the LUMI GPU partition (LUMI-G). The strong scaling is obtained as the ratio of the elapsed time for one node ($T_1$) to the elapsed time ($T$) for a grid of about 1.1 billion points. Weak scaling is obtained as the ratio of elapsed time ($T$) to elapsed time for one node ($T_1$), with each node holding a grid of about 1.1 billion points. Each node consists of 4 AMD MI250x GPUs or 8 GCDs with a total of 512 GiB of memory.}
    \label{fig:lumi_scalability}
\end{figure}

Each LUMI node on the GPU partition consists of eight GCDs (four GPUs in total). HIPFort is compiled with hipfc/hipcc (based on ROCm 5.0.2) and MPICH, which supports ROCm-aware MPI exchanges. In addition, all nodes are equipped with a Slingshot-11 interconnect, with one of the interfaces connected directly to each GPU. Figures~\ref{fig:lumi_scalability}a and \ref{fig:lumi_scalability}b illustrate the strong and weak scalability performance of the HIPFort backend. For the strong scalability we use a grid consisting of 1.1 billion points, which for the reference case of a single node corresponds to a memory utilisation of about 91\% out of the available 512 GiB (eight GCDs). For weak scalability, we use the same grid for each node. 

The strong scalability plot (figure~\ref{fig:lumi_scalability}a) shows an initial superlinearity up to 16 nodes for all cases. Beyond 16 nodes, the profiles begin to diverge from the linear case. Considering the strong scaling efficiency, which is the percentage ratio of the achieved speedup to the ideal speedup, we achieve more than 80\% up to 32 nodes. Asynchronous communication has a large advantage over the synchronous case for the central scheme, while it is slightly better for WENO. As for the weak scalability performance ( figure~\ref{fig:lumi_scalability}b), we see a significant weak scaling efficiency (percentage ratio of ideal to achieved speedup) greater than 97\% in all cases. Asynchronous communication has the upper hand again in both the central and WENO cases as the number of nodes increases, with efficiencies above 98\%.  

\begin{figure}[H]
    \centering
    \includegraphics{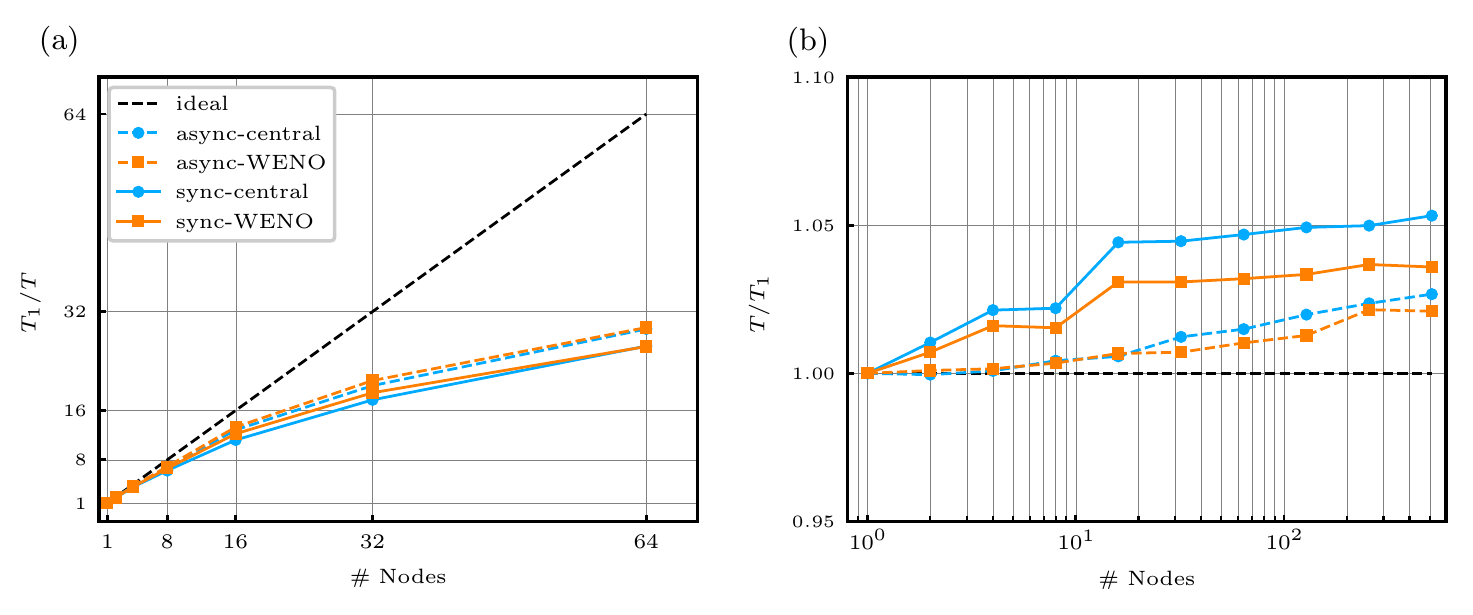}
    \caption{Strong \textit{(a)} and weak \textit{(b)} scalability plots for STREAmS-2 in synchronous and asynchronous modes for central and WENO flux schemes on Leonardo GPU partition. The strong scaling is obtained as the ratio of the elapsed time for one node ($T_1$) to the elapsed time ($T$) for a grid of about 0.6 billion points. Weak scaling is obtained as the ratio of elapsed time ($T$) to elapsed time for one node ($T_1$), with each node holding a grid of about 0.6 billion points. Each node consists of 4 NVIDIA A100 GPUs with a total of 256 GiB of memory.}
    \label{fig:leo_scal}
\end{figure}

Each Leonardo node on the GPU partition consists of four GPUs. The backend is compiled using the NVIDIA HPC-SDK compiler (version 23.1) with support for CUDA-aware MPI transfers. In addition, all nodes are interconnected using the Quad-rail NVIDIA HDR100 Infiniband. Figures~\ref{fig:leo_scal}a and \ref{fig:leo_scal}b illustrate the strong and weak scalability performance of the CUDA Fortran backend. We use the same strategy to perform the scalability evaluation, similar to the LUMI case, but we consider a grid of 0.6 billion points, which for the reference case of a single node corresponds to a memory usage of about 90\% of the available 256 GiB (four GPUs).

On the strong scalability plot based on Figure~\ref{fig:leo_scal}a, we see that the achieved speed deviates from the ideal line at around 16 nodes. The strong scaling efficiencies at 32 nodes drop to around 55-65\% for all cases. However, the weak scaling results are excellent, with efficiencies for the asynchronous case greater than 98\%. Again, the asynchronous mode performs better for both strong and weak scaling.

Comparing figures~\ref{fig:lumi_scalability}a and \ref{fig:leo_scal}a we see that STREAmS-2 has better strong scalability performance for the HIPFort backend on LUMI. In particular, superlinearity only occurs in this case. This result can be attributed to the different behaviour of the GPUs when varying the number of grid points processed, as shown in figure~\ref{fig:grind_time}. We can see that as the number of points increases, the grind time for the NVIDIA GPU continues to decrease. For the AMD counterpart, however, it initially decreases, but then increases again. In particular, the optimal grind time for the NVIDIA card occurs when the number of grid points per GPU is at its highest, but for the AMD GPU it occurs when it is relatively lower. When the number of GPUs is increased for strong scaling, the effect of reducing the number of grid points per GPU is superimposed on the MPI communication overhead. For the NVIDIA case, the two effects are always additive, resulting in a progressively decreasing scaling efficiency. For the AMD case, the situation is more complex: in the first part of the scaling, the performance improvement related to the grid reduction seems to outweigh the relative increase in communication overhead, while in the second part, the grid reduction effect and the communication overhead have the same effect, leading to the final performance degradation.

The weak scalability comparison, based on figures~\ref{fig:lumi_scalability}b and \ref{fig:leo_scal}b, shows impressive efficiencies for both synchronous and asynchronous modes. However, the asynchronous mode always reduces the efficiency loss compared to the synchronous mode. For the CUDA Fortran backend, at the highest number of nodes, this reduction is around 1.4\% for WENO and 2.5\% for central. Similarly, for the HIPFort backend, this reduction is less pronounced but still measurable at around 1\%. On the other hand, using asynchronous mode has a greater advantage when using smaller grid sizes per GPU, as evidenced by the strong scalability plots for both backends. 

The results have important implications for the capabilities of the EuroHPC JU pre-exascale clusters. From figure~\ref{fig:lumi_scalability}b, the number of points based on 256 nodes is about 301 billion. Based on this speedup trend, we can efficiently run cases on the order of 1 trillion points, which will certainly help to study supersonic/hypersonic flow physics on an unprecedented scale.  

\section{Conclusions} \label{sec:discussion}

With the aim of taking full advantage of the powerful modern HPC GPU architectures, we present an enhanced version of STREAmS-2 for the study of compressible turbulent wall-bounded flows. The GPUs 
under scrutiny are based on the NVIDIA and AMD architectures. The framework and strategies outlined in this paper should allow STREAmS-2 to tackle complex flow problems through high fidelity simulations of an unprecedented level, usually considered implausible. We have described a development model built on a multi-equation, multi-backend framework, where the primary code version is based on CUDA Fortran, allowing efficient use of NVIDIA GPUs while maintaining good readability, which is crucial especially when developing new code features. On the other hand, a Python tool was developed to convert the CUDA Fortran code into traditional CPU and HIPFort paradigms, the latter suitable for AMD GPUs. It was essential to follow an object-based design and strict programming guidelines to ensure a clean conversion process. 

The single-GPU evaluations show the impressive performance of GPUs compared to traditional CPUs. In addition, the detailed investigations allow interesting comparisons to be made between the two GPUs currently at the heart of many EuroHPC systems, the NVIDIA A100 and the AMD MI250x. In general, it was found that the gap between peak performance and actual STREAmS-2 performance is generally smaller for NVIDIA cards. Roofline analysis allows us to partially trace the reason for this behaviour and shows that, somewhat unexpectedly, the computational intensity of the core kernels is significantly different for the two architectures. Additional single-GPU comparisons are performed to assess the impact of grid size, number of parallelised loops, thread masking and thread divergence. The results obtained can be of interest to users in configuring their execution setup and to developers in efficiently programming new algorithms.

The parallel performance has been evaluated in terms of strong and weak scaling, taking into account synchronous and asynchronous communication patterns and running on the two largest EuroHPC clusters, namely LUMI (CSC) and Leonardo (Cineca). The strong scalability shows more than 80\% efficiency up to 16 nodes for Leonardo and up to 32 nodes for LUMI. Weak scalability shows an impressive efficiency of over 95\% up to the maximum number of nodes tested (256 for LUMI and 512 for Leonardo). Enabling the asynchronous pattern for communication-computation overlap allows significant improvements. Overall, the final performances on these clusters allow us to understand that impressive computational grids can be adopted, potentially approaching trillions of grid points. Thus, the physical and engineering problems that will be addressed by STREAmS-2 have the full potential to cover new areas and applications of compressible fluid dynamics. 

\section*{Acknowledgments}
Funded by the European Union. This work has received funding from the European High Performance Computing Joint Undertaking (EuroHPC JU) and Germany, Italy, Slovenia, Spain, Sweden, and France under grant agreement No 101092621. We also acknowledge Cineca (ISCRA project \textit{IscrB\_SCFRG}, LEAP project \textit{DISCOVERER}), CSC (Development project \textit{EHPC-DEV-2021D04-131}, Benchmarking project \textit{EHPC-BEN-2023B02-008}) and HLRS Supercomputing Centers for computational resources and support in carrying out the tests. Matteo Bernardini has also been supported by the grant agreement No A0375-2020-36614, project GREEN H2 CFD, POR FESR LAZIO 2014-2020.

\bibliography{references}

\begin{thebibliography}{10}
\expandafter\ifx\csname url\endcsname\relax
  \def\url#1{\texttt{#1}}\fi
\expandafter\ifx\csname urlprefix\endcsname\relax\def\urlprefix{URL }\fi
\expandafter\ifx\csname href\endcsname\relax
  \def\href#1#2{#2} \def\path#1{#1}\fi

\bibitem{top500}
{TOP}500, 2022, \url{https://www.top500.org/lists/top500/2022/11/}, accessed:
  25 February 2023.

\bibitem{cuda}
{CUDA}, 2023, \url{https://docs.nvidia.com/cuda/cuda-c-programming-guide/},
  accessed: 25 February 2023.

\bibitem{zhu2018afid}
X.~Zhu, E.~Phillips, V.~Spandan, J.~Donners, G.~Ruetsch, J.~Romero,
  R.~Ostilla-M{\'o}nico, Y.~Yang, D.~Lohse, R.~Verzicco, M.~Fatica, R.~Stevens,
  A{F}i{D}-{GPU}: a versatile {N}avier--{S}tokes solver for wall-bounded
  turbulent flows on {GPU} clusters, Comput. Phys. Commun. 229 (2018) 199--210.

\bibitem{costa2021gpu}
P.~Costa, E.~Phillips, L.~Brandt, M.~Fatica, {GPU} acceleration of {C}a{NS} for
  massively-parallel direct numerical simulations of canonical fluid flows,
  Comput. \& Math. Appl. (2020).

\bibitem{cuda-fortran}
{CUDA F}ortran, 2023,
  \url{https://docs.nvidia.com/hpc-sdk/compilers/cuda-fortran-prog-guide/},
  accessed: 25 February 2023.

\bibitem{openacc}
{O}pen{ACC}, 2023, \url{https://docs.nvidia.com/hpc-sdk/compilers/openacc-gs/},
  accessed: 25 February 2023.

\bibitem{witherden2014pyfr}
F.~Witherden, A.~Farrington, P.~Vincent, {PyFR}: {A}n open source framework for
  solving advection--diffusion type problems on streaming architectures using
  the flux reconstruction approach, Comput. Phys. Commun. 185~(11) (2014)
  3028--3040.

\bibitem{romero2020zefr}
J.~Romero, J.~Crabill, J.~Watkins, F.~Witherden, A.~Jameson, {ZEFR}: A
  {GPU}-accelerated high-order solver for compressible viscous flows using the
  flux reconstruction method, Comput. Phys. Commun. 250 (2020) 107169.

\bibitem{de2023gpu}
F.~De~Vanna, F.~Avanzi, M.~Cogo, S.~Sandrin, M.~Bettencourt, F.~Picano,
  E.~Benini, {URANOS}: A {GPU} accelerated navier-stokes solver for
  compressible wall-bounded flows, Comput. Phys. Commun. (2023) 108717.

\bibitem{bres_22}
G.~A. Bres, S.~T. Bose, C.~B. Ivey, M.~Emory, F.~Ham, {GPU}-accelerated
  large-eddy simulations of supersonic jets from twin rectangular nozzle
  (2022).

\bibitem{kuznik2010lbm}
F.~Kuznik, C.~Obrecht, G.~Rusaouen, J.-J. Roux, {LBM} based flow simulation
  using {GPU} computing processor, Comput. \& Math. Appl. 59~(7) (2010)
  2380--2392.

\bibitem{zhang2020novel}
J.-L. Zhang, H.-Q. Chen, S.-G. Xu, H.-Q. Gao, A novel {GPU}-parallelized
  meshless method for solving compressible turbulent flows, Comput. \& Math.
  Appl. 80~(12) (2020) 2738--2763.

\bibitem{jansson2021neko}
N.~Jansson, M.~Karp, A.~Podobas, S.~Markidis, P.~Schlatter, {N}eko: {A}
  {M}odern, {P}ortable, and {S}calable {F}ramework for {H}igh-{F}idelity
  {C}omputational {F}luid {D}ynamics, arXiv preprint arXiv:2107.01243 (2021).

\bibitem{hip}
{HIP}: {C}++ {H}eterogeneous-{C}ompute {I}nterface for {P}ortability, 2023,
  \url{https://github.com/ROCm-Developer-Tools/HIP/}, accessed: 25 February
  2023.

\bibitem{opencl}
{O}pen{CL}, 2023,
  \url{https://registry.khronos.org/OpenCL/specs/3.0-unified/pdf/OpenCL_API.pdf},
  accessed: 25 February 2023.

\bibitem{salvadore2013gpu}
F.~Salvadore, M.~Bernardini, M.~Botti, {GPU} accelerated flow solver for direct
  numerical simulation of turbulent flows, J. Comput. Phys. 235 (2013)
  129--142.

\bibitem{bernardini2021streams}
M.~Bernardini, D.~Modesti, F.~Salvadore, S.~Pirozzoli, {STREAmS}: a
  high-fidelity accelerated solver for direct numerical simulation of
  compressible turbulent flows, Comput. Phys. Commun. 263 (2021) 107906.

\bibitem{bernardini2023streams}
M.~Bernardini, D.~Modesti, F.~Salvadore, S.~Sathyanarayana, G.~Della~Posta,
  S.~Pirozzoli, {STREAmS}-2.0: {S}upersonic turbulent accelerated
  {N}avier-{S}tokes solver version 2.0, Comput. Phys. Commun. (2023) 108644.

\bibitem{pirozzoli_bernardini_2011}
S.~Pirozzoli, M.~Bernardini, Turbulence in supersonic boundary layers at
  moderate {R}eynolds number, J. Fluid Mech. 688 (2011) 120--168.

\bibitem{pirozzoli_probing}
S.~Pirozzoli, M.~Bernardini, Probing high-{R}eynolds-number effects in
  numerical boundary layers, Phys.~Fluids 25~(2) (2013) 021704.

\bibitem{cogo_salvadore_picano_bernardini_2022}
M.~Cogo, F.~Salvadore, F.~Picano, M.~Bernardini, Direct numerical simulation of
  supersonic and hypersonic turbulent boundary layers at moderate-high
  {R}eynolds numbers and isothermal wall condition, J. Fluid Mech. 945 (2022)
  A30.

\bibitem{bernardini_della_posta_salvadore_martelli_2023}
M.~Bernardini, G.~Della~Posta, F.~Salvadore, E.~Martelli, Unsteadiness
  characterisation of shock wave/turbulent boundary-layer interaction at
  moderate {R}eynolds number, J. Fluid Mech. 954 (2023) A43.

\bibitem{modesti2016reynolds}
D.~Modesti, S.~Pirozzoli, {R}eynolds and {M}ach number effects in compressible
  turbulent channel flow, Int. J. Heat and Fluid Flow 59 (2016) 33--49.

\bibitem{bernardini2012compressibility}
M.~Bernardini, S.~Pirozzoli, P.~Orlandi, Compressibility effects on
  roughness-induced boundary layer transition, Int. J. Heat and Fluid Flow 35
  (2012) 45--51.

\bibitem{roughness_bernardini}
M.~Bernardini, S.~Pirozzoli, P.~Orlandi, S.~Lele, Parameterization of
  {B}oundary-{L}ayer {T}ransition {I}nduced by {I}solated {R}oughness
  {E}lements, AIAA J. 52~(10) (2014) 2261--2269.

\bibitem{modesti_sathyanarayana_salvadore_bernardini_2022}
D.~Modesti, S.~Sathyanarayana, F.~Salvadore, M.~Bernardini, Direct numerical
  simulation of supersonic turbulent flows over rough surfaces, J. Fluid Mech.
  942 (2022) A44.

\bibitem{pirozzoli_10}
S.~Pirozzoli, Generalized conservative approximations of split convective
  derivative operators, J. Comput. Phys. 229~(19) (2010) 7180--7190.

\bibitem{tamaki_22}
Y.~Tamaki, Y.~Kuya, S.~Kawai, Comprehensive analysis of entropy conservation
  property of non-dissipative schemes for compressible flows: {KEEP} scheme
  redefined, J. Comput. Phys. 468 (2022) 111494.

\bibitem{mcbride_02}
B.~McBride, {NASA} {G}lenn coefficients for calculating thermodynamic
  properties of individual species, National Aeronautics and Space
  Administration, John H. Glenn Research Center, 2002.

\bibitem{oneapi}
one{API}, 2023, \url{https://www.oneapi.io/}, accessed: 31 March 2023.

\bibitem{openmp}
{O}pen{MP}, 2023,
  \url{https://www.openmp.org/wp-content/uploads/OpenMP-API-Specification-5-2.pdf/},
  accessed: 25 February 2023.

\bibitem{kokkos}
kokkos, 2023, \url{https://github.com/kokkos/kokkos}, accessed: 31 March 2023.

\bibitem{legion}
{L}egion, 2023, \url{https://legion.stanford.edu/}, accessed: 31 March 2023.

\bibitem{opensycl}
{O}pen{SYCL}, 2023, \url{https://github.com/OpenSYCL/OpenSYCL}, accessed: 31
  March 2023.

\bibitem{alpaka}
alpaka, 2023, \url{https://github.com/alpaka-group/alpaka}, accessed: 31 March
  2023.

\bibitem{raja}
{RAJA}, 2023,
  \url{https://computing.llnl.gov/projects/raja-managing-application-portability-next-generation-platforms},
  accessed: 31 March 2023.

\bibitem{hipify}
{HIPIFY}: {C}onvert {CUDA} to {P}ortable {C}++ code, 2023,
  \url{https://github.com/ROCm-Developer-Tools/HIPIFY}, accessed: 25 February
  2023.

\bibitem{HIPFort}
{HIPF}ort, 2023, \url{https://github.com/ROCmSoftwarePlatform/hipfort/},
  accessed: 25 February 2023.

\bibitem{hipcub}
hip{CUB}, 2023, \url{https://github.com/ROCmSoftwarePlatform/hipCUB/},
  accessed: 25 February 2023.

\bibitem{gpufort}
{GPUFORT}, 2023, \url{https://github.com/ROCmSoftwarePlatform/gpufort/},
  accessed: 25 February 2023.

\bibitem{eurohpc}
{EUROHPC JU}, 2023,
  \url{https://eurohpc-ju.europa.eu/about/our-supercomputers_en}, accessed: 31
  March 2023.

\bibitem{yang2020hierarchical}
C.~Yang, T.~Kurth, S.~Williams, Hierarchical {R}oofline analysis for {GPU}s:
  {A}ccelerating performance optimization for the {NERSC}-9 {P}erlmutter
  system, Concurrency and Computation: Practice and Experience 32~(20) (2020)
  e5547.

\end{thebibliography}

\end{document}